\documentclass[11pt,a4paper]{article}
\usepackage{jheppub}
\usepackage[numbers]{natbib}
\usepackage{dsfont}
\usepackage{xcolor}
\usepackage{empheq}
\usepackage{graphicx}
\usepackage{dcolumn}
\usepackage{bm}

	\newcommand{\beq}{\begin{equation}}   
		\newcommand{\eeq}{\end{equation}}
	\newcommand{\beqn}{\begin{eqnarray}}   
		\newcommand{\eeqn}{\end{eqnarray}}
	\newcommand{\pt}{\partial}

	\newcommand{\ntwo}{{\mathcal N}=2}
	\newcommand{\nfour}{{\mathcal N}=4}
	\newcommand{\none}{{\mathcal N}=1}

\title{\boldmath Peculiarities of beta functions in sigma models}

\author[a,1]{Oleksandr Gamayun,\note{Corresponding author.}}
\author[b,c]{Andrei Losev}
\author[d]{Mikhail Shifman}

\affiliation[a]{London Institute for Mathematical Sciences, Royal Institution,\\ 21 Albemarle St,
London W1S 4BS, UK}
\affiliation[b]{Wu Wen-Tsun Key Lab of Mathematics, Chinese Academy of Sciences}
\affiliation[c]{National Research University Higher School of Economics, Moscow, Russia}
\affiliation[d]{William I. Fine Theoretical Physics Institute, University of Minnesota, \\ Minneapolis, MN 55455, USA}

\emailAdd{og@lims.ac.uk}
\emailAdd{aslosev2@yandex.ru}
\emailAdd{shifman@umn.edu}

\preprint{FTPI-MINN-23-11, UMN-TH-4218/23}	

\abstract{In this paper we consider perturbation theory in generic two-dimensional sigma models in the so-called first-order formalism, using
the coordinate regularization approach. Our goal is to analyze the first-order formalism in application to $\beta$ functions and compare its results  
with the standard geometric calculations. 
Already in the second loop, we observe deviations from the geometric results that cannot be explained 
by the regularization/renormalization scheme choices. Moreover, in certain cases the first-order calculations produce results that are not symmetric
under the classical diffeomorphisms of the target space. Although we could not present the full solution to this remarkable phenomenon, we found some indirect arguments indicating that  an anomaly similar to that established in supersymmetric Yang-Mills theory
might manifest itself starting from the second loop.  
We  discuss why the difference between two answers might be an infrared effect, similar to that in $\beta$ functions in supersymmetric Yang-Mills theories.

In addition to the generic K\"ahler target spaces  we discuss in detail the so-called  Lie-algebraic sigma models. In particular, this is the case when the perturbed field  $G^{i\bar j}$ is a product of the holomorphic and antiholomorphic currents satisfying two-dimensional current algebra. 
}

\begin{document} 
\maketitle
\flushbottom 

\section{Introduction}

Sigma models on K\"ahlerian and super-K\"ahlerian target spaces play an important role in strings and various aspects of quantum field theory --
from mathematics to phenomenology. If supersymmetrized in two dimensions, they have extendended supersymmetries, $\ntwo$ or $\nfour$ \footnote{
 In addition to ${\mathcal N}$=(2,2), the K\"ahler spaces allow for chiral spersymmetrizations  ${\mathcal N}$=(0,2) or (2,0).
}.
In generic  K\"ahlerian models with non-Einstein target space each order of perturbation theory brings in new geometric structures, (see e.g. \cite{Grah,Foakes1987,Foakes1988,Ketov2000,Naka}), say, in the first- and second-loop orders we have
\begin{equation}
{\mathcal L}_{\rm eff} =\left\{  G_{i\bar j} +\left(\alpha' R_{i\bar j} + \frac{(\alpha')^2}{2} R_{i\bar k\ell \bar m}\, R_{\bar j}^{\,\,\bar k\ell \bar m}\right)L+...
\right\}\partial_\mu\varphi^i \partial_\mu\bar\varphi^{\,\bar j}
\label{one}
\end{equation}
where
$L$ in Eq. (\ref{one}) is defined as
\begin{equation}\label{log}
L=-\log\frac{M_{\rm uv}}{ \mu}\,.
\end{equation}
Neither $R_{i\bar j}$ nor $R_{i\bar k\ell \bar m}\, R_{\bar j}^{\,\,\bar k\ell \bar m} $ reduce to $G_{i\bar{j}}$ in the general case. 

In this paper, we first consider renormalization aspects of the so-called first order formalism in generic sigma models \cite{Losev2006} (for brevity referred to
as the first-order sigma models). 
We treat these models as deformations of the conformal field theories of the simple bosonic ``$\beta$-$\gamma$'' systems by using the so-called coordinate approach to compute the beta function. We consider correlation functions in the real space (opposite to the traditional momentum) and employ the point-splitting regularization.   
This construction is {\em a priori} expected to produce beta functions of  a certain self-repeating pattern of the (differential) polynomial type for the inverse metric $G^{i\bar j}$.  
Comparing with Eq. \eqref{one} we see that the expected pattern does emerge at one loop (see below). Although generalizing our analysis 
to higher loops we continue to observe the polynomial pattern, the $\beta$ function obtained in this way does {\em not} coincide
with the result of the standard geometric calculation presented in (\ref{one}) starting from the second loop   \cite{Grah,Foakes1987,Foakes1988,Tsey}. Three-loop contribution
is RG-scheme dependent -- therefore one might blame the difference in the renormalization schemes for the disagreement.\footnote{ It was calculated by Foakes and Mohammedi, and Graham in a
	particular scheme  \cite{Grah,Foakes1987,Foakes1988}. }

However, the disagreement  we observe appears already in the second loop. This leads us to the hypothesis that starting from the second loop a certain anomaly
manifests itself, perhaps, similar to that established in supersymmetric Yang-Mills theory \cite{NSZV,NSZV1,CS,AH}. Some indirect arguments 
in favor of this hypothesis will be given in Sec. \ref{hyp}.

Besides the generic K\"ahler target spaces  we discuss in detail the so-called  Lie-algebraic sigma models\footnote{The  quantum-mechanical Hamiltonians with the Lie-algebraic structure led to the discovery of quasi-exactly solvable systems, see reviews \cite{tur,shif}.}. In paricular, this is the case when the perturbed field  $G^{i\bar j}$ is a product of the holomorphic and antiholomorphic currents satisfying two-dimensional current algebra. 
We obtain exact formulas for the one and two-loop beta functions using only algebraic data and observe the {\em full correspondence} with the known results for current-current deformations of the conformal field theories \cite{Kutasov1989,PhysRevLett.86.4753,Ludwig2003} (but not with the geometric formulas).

The paper is organized as follows. 
In Sec. \eqref{sec:beta1} we introduce first-order sigma model, outline the two-loop computation of the $\beta$ function and discuss why the answer is different from the one in Eq. \eqref{one}. 
In Sec. \eqref{sec:currents} we discuss an important example of the current-current perturbation, which we also dub as {\em Lie-algebraic sigma models}, and show why the answer in Sec. \eqref{sec:beta1} makes sense from the algebraic point of view.   In Sec. \eqref{hyp} we indicate that discrepancies
between the``geometric"  and ``algebraic" $\beta$ functions can be an IR effect. Illuminating
instanton calculations are presented in Sec. \eqref{sec:insta}.

	\section{Coordinate \boldmath{$\beta$} function}\label{sec:beta1}

 \subsection{Model and derivation}
	
The first order sigma model proposed in \cite{Losev2006} and further developed in \cite{Zeitlin2006,Zeitlin2008,Zeitlin2009}, 
was initially introduced as a tool to address singular background in string theory. 
More specifically, the bare action is described by the following conformal field theory 
\begin{equation}\label{S0}
	S_0 = \frac{1}{\alpha'}\int \frac{d^2z}{\pi}  \left(
	p_a \bar{\partial}\varphi^a + \bar{p}_{\bar{a}}\partial \bar\varphi^{\bar{a}}
	\right).
\end{equation}
Here $d^2z = dx dy= idz\wedge d\bar{z}/2$, the index $a$ runs from $1$ to $D/2$ (here we have assumed that the target space is a complex manifold with local coordinates $\varphi^a$ and $\bar\varphi^{\bar{a}}$). The fields $p_a$, $p_{\bar{a}}$ are $(1,0)$  and $(0,1)$ forms on $\Sigma$ correspondingly, and the fields $\varphi^a$, $\varphi^{\bar{a}}$ are scalars. 
For our purposes, it will be enough to consider $\Sigma = \mathds{C}P^1$. For more precise definitions and discussion about the choice of complex structure see \cite{nekrasov2005lectures}. To simplify notations in what follows we put $\alpha'=1$.

The action \eqref{S0} is equivalent to the  following operator product expansions,
\begin{equation}\label{OPE1}
	p_a(z) \varphi^b(w) = \frac{\delta_a^b}{z-w} + {\rm reg},\qquad
	\bar{p}_{\bar{a}}(\bar{z}) \bar{\varphi}^{\bar{b}}(\bar{w}) = \frac{\delta_{\bar{a}}^{\bar{b}}}{\bar{z}-\bar{w}} + {\rm reg}.
\end{equation}
To turn to the non-singular sigma models one considers perturbation of the action (\ref{S0}) by four possible classically marginal terms 
\begin{equation}\label{FullS}
	S = S_0 - \int \frac{d^2z}{\pi} (O_G+ O_{
		\mu} + O_{\bar \mu} + O_b)
\end{equation}
 where 
\begin{equation}\label{deform}
	O_G = G^{a\bar{a}}p_a\bar{p}_{\bar{a}},\qquad  	O_\mu =p_a\mu^{a}_{\bar{a}}\bar{\partial}\bar{\varphi}^{\bar{a}}
	,\qquad 	O_{\bar{\mu}} =\bar{p}_{\bar{a}}\bar{\mu}_{a}^{\bar{a}}\partial \varphi^{a},\qquad O_b = b_{a\bar{a}}\partial \varphi^{a}\bar{\partial} \bar{\varphi}^{\bar{a}}.
\end{equation} 
Assuming that $G^{a\bar{a}}$ is not degenerate one can integrate over $p$ and $\bar{p}$ and obtain the second-order sigma model with the metric and Kalb-Ramond fields parametrized in terms of the data $G$, $b$, $\mu$ and $\bar{\mu}$ (see \cite{Losev2006}). 
Besides, an additional field of a dilaton appears due to the anomaly in the path integral measure. 

For operators \eqref{deform} to be primary, in the first order of perturbation theory one has to impose transversality condition 
\begin{equation}\label{Trans}
    \partial_a G^{a\bar{a}} =  \partial_{\bar{a}} G^{a\bar{a}} =\partial_a \mu^{a}_{\bar{a}} = \partial_{\bar{a}}\bar{\mu}_{a}^{\bar{a}} = 0. 
\end{equation}
This in particular allows one to avoid possible self-contractions of $p$ and $\varphi$ inside operators. 
In what follows, we will not insist on the transversality condition and presume that self-contractions are subtracted, in other words the operators are normal-ordered. 

In this paper we focus mainly on the  case when only $O_G$ perturbations are present -- then the metric and Kalb-Ramond fields are, respectively, symmetric and antisymmetric forms of inverse of the bivector $G^{a\bar{a}}$, which we denote by $G_{a\bar{a}}$.  The dilaton field $\Phi$ is proportional to the logarithm of the determinant $$\Phi \sim \log \det G\,.$$  Note that for K\"ahlerian metrics with constant determinant the effects due to the  Kalb-Ramond fields and dilatons are irrelevant. 

Here again, we stress that the first order system makes sense even when the inverse of $G^{a\bar{a}}$ does not exist. Our point of view is that 
the {\it original} sigma model is represented by Lagrangian 
\eqref{FullS}. In this paper we focus on making sense of \eqref{FullS} as a quantum field theory. 

First, let us focus on the case when $b_{a\bar{a}} =\bar{\mu}_{a}^{\bar{a}}=\mu^{a}_{\bar{a}} = 0 $, i.e. on the action 
	\begin{equation}
		S = \int \frac{d^2z}{\pi}  \left(
		p_a \bar{\partial}\varphi^a + \bar{p}_{\bar{a}}\partial \bar\varphi^{\bar{a}} - G^{a\bar{a}}(\varphi,\bar{\varphi}) p_a\bar{p}_{\bar{a}}	
		\right).
	\end{equation}
We can regard $G$ as a perturbation for the conformal field theory given by $S_0$  in Eq. \eqref{S0}.
Let us demonstrate the necessity of renormalization using the example of the perturbative calculation of the following two-point function, 
		\begin{equation}
		C_2(x,y) =  \Big\langle B(x)O_g(y) \Big\rangle_G \equiv \Big\langle B(x)O_g(y)\exp\left(\int_{|z|<R}\frac{d^2z}{\pi} O_G\right)\Big\rangle_{0}   .
		\label{eight}
	\end{equation}
	Here $$B(x) \equiv b_{c\bar{c}}\partial\varphi^c\bar{\partial}\bar\varphi^{\bar{c}}$$ is a composite field which is denoted differently from $O_b$ to emphasize the fact that the field $b_{a\bar{a}}$ is 
	an observable  (and not the one in the action). By $O_g$ we denote $g^{a\bar{a}}p_a\bar{p}_{\bar{a}}$ again to emphasize different role of the observable field $g^{a\bar{a}}$ as opposed to  to $G^{a\bar{a}}$ 
	in the exponential. All observables are considered to be normal ordered, so we don't have to worry about self-contractions.
	The effect of non-zero $G$ is included order by order in the perturbation theory which is symbolized by the last equality in  (\ref{eight}).  The subscript there reflects that computations are carried out
	in the free theory \eqref{S0}.
	Here we have also introduced an IR cut-off $R$ and assume that $|x|\ll R$, $y\ll R$.
	In the zeroth order using OPE \eqref{OPE1} we conclude
	\begin{equation}
		C_2^{(0)} (x,y)= \frac{g^{a\bar{a}}b_{a\bar{a}}}{|x-y|^4} .
	\end{equation}
	
	At this point, we have to clarify what we mean by correlation functions as functions of the target space fields.  
 Strictly speaking instead of $g^{a\bar{a}}b_{a\bar{a}}$ one has to consider  an integral 
	\begin{equation}\label{gb}
		\int \,g^{a\bar{a}} (\varphi^{(0)}) b_{a\bar{a}} (\varphi^{(0)}) \, d^{D}\varphi^{(0)} 
	\end{equation}
	over the  modes $\varphi^{(0)}$ of the $\varphi$-fields. We assume that the target space is compact, or that the observables $g^{a\bar{a}}$, $b_{a\bar{a}}$ 
	are functions rapidly vanishing at the ``space-time infinity,'' so that all integrals are convergent.  
 
 \textit{We  assume this prescription for all the correlation functions in this paper. }
 
 In particular, this will allow us to integrate by parts in some of the examples below. 
	Note also that we use Wick's theorem to contract derivatives of the field, which would, otherwise, vanish after the integration.

	The next order is given by an integral of the three-point function 
	\begin{equation}
		C_2^{(1)} (x,y)=  \int\limits_{|z|<R} \frac{d^2z}{\pi} \Big\langle B(x)O_g(y)O_G(z)\Big\rangle_{0} .
	\end{equation}
	In this case the computation of the correlation function is a bit more complicated, but still can be performed explicitly. Namely, after contracting the $p$ field with the derivatives we arrive at 
	\begin{equation}
		\Big\langle B(x)O_g(y)O_G(z)\Big\rangle_{0}  = 
		\frac{\langle b_{a\bar{a}}(x) g^{a\bar{a}}(y) G^{c\bar{c}}p_c\bar{p}_{\bar{c}}(z)\rangle}{|x-y|^4} + 
		\frac{\langle b_{a\bar{a}}(x) g^{a\bar{b}}\bar{p}_{\bar{b}}(y) G^{c\bar{a}}p_c(z)\rangle}{(x-y)^2 (\bar{x}-\bar{z})^2}+ \left(
		\begin{array}{c}
			y \leftrightarrow z \\ 
			G \leftrightarrow g
		\end{array}
		\right).
	\end{equation}
	Contracting the remaining $p$ field we obtain 
	\begin{equation}
		\frac{\langle b_{a\bar{a}}(x) g^{a\bar{a}}(y) G^{c\bar{c}}p_c\bar{p}_{\bar{c}}(z)\rangle}{|x-y|^4}= 
		\frac{G^{c\bar{c}}\partial_c\partial_{\bar{c}} b_{a\bar{a}}(x) g^{a\bar{a}}(y) }{|x-y|^4|z-x|^2} + 
		\frac{G^{c\bar{c}}\partial_c b_{a\bar{a}}(x)\partial_{\bar{c}}g^{a\bar{a}}(y) }{|x-y|^4(z-x)(\bar{z}-\bar{y})} 
		+ \left(
		\begin{array}{c}
			y \leftrightarrow x \\ 
			g \leftrightarrow b
		\end{array}
		\right),
	\end{equation}
	\begin{equation}
		\langle b_{a\bar{a}}(x) g^{a\bar{b}}\bar{p}_{\bar{b}}(y) G^{c\bar{a}}p_c(z)\rangle  = 
		\frac{\partial_{\bar{b}}\partial_c b_{a\bar{a}}g^{a\bar{b}}G^{c\bar{a}}}{(\bar{y}-\bar{x})(z-x)} 
		+ \frac{\partial_{\bar{b}}b_{a\bar{a}}\partial_c g^{a\bar{b}}G^{c\bar{a}}}{(\bar{y}-\bar{x})(z-y)}  +
		\frac{b_{a\bar{a}}\partial_c g^{a\bar{b}}\partial_{\bar{b}}G^{c\bar{a}}}{(\bar{y}-\bar{z})(z-y)}  
		+ \frac{\partial_{c}b_{a\bar{a}} g^{a\bar{b}}\partial_{\bar{b}}G^{c\bar{a}}}{(\bar{y}-\bar{z})(z-x)} 
	\end{equation}
	Assembling everything together we obtain an expression for the correlator that does not look like a typical three point function in CFT. 
	This happened because  we relaxed the transversality condition, which on this level is equivalent to the statement that operators are primary. 
	Anyhow, from the obtained expression one might expect possible divergences when the integration comes close to $x$ or $y$, therefore we add 
	UV regularization prescription that small disks of the size  $\epsilon$ are excluded from the  integration domain, namely $|z-x|<\epsilon$, $|z-y|<\epsilon$. 
	Performing integration with such a prescription we obtain 
	\begin{equation}\label{c21}
		C_2^{(1)}(x,y) = \frac{\log \frac{|x-y|^2}{\epsilon^2}}{|x-y|^4} \Big(
		g^{a\bar{a}}\boldsymbol{[}b,G\boldsymbol{]}_{a\bar{a}} + b_{a\bar{a}} \boldsymbol{[}g,G\boldsymbol{]}^{a\bar{a}} 
		\Big) + \delta C_2^{(1)}(x,y) 
	\end{equation}
	where $ \delta C_2^{(1)}$ is present only due to effects of the fields being non-primary (non-tranversal in the sense of Eq. \eqref{Trans}), 
\begin{multline}
\delta C_2^{(1)}(x,y) = \frac{G^{c\bar{c}}\partial_c\partial_{\bar{c}}(b_{a\bar{a}}g^{a\bar{a}})}{|x-y|^4} \log \frac{R^2}{|x-y|^2}
+ \frac{4}{\epsilon^2} \frac{g^{c\bar{c}}\partial_c\partial_{\bar{c}} (b_{a\bar{a}}G^{a\bar{a}})}{|x-y|^2}  \\
+	\frac{\partial_c(b_{a\bar{a}}G^{a\bar{c}}) \partial_{\bar{c}} g^{c\bar{a}}+ \partial_c g^{a\bar{c}}\partial_{\bar{c}}(b_{a\bar{a}}G^{c\bar{a}}) -g^{c\bar{c}}\partial_{\bar{c}}(b_{a\bar{a}}\partial_cG^{a\bar{a}})-g^{c\bar{c}}\partial_{c} (b_{a\bar{a}}\partial_{\bar{c}}G^{a\bar{a}})}{|x-y|^4}.
	\end{multline}
Notice that if trasversality condition \eqref{Trans} is satisfied then $\delta C_2^{(1)}$ vanishes after integration by parts in the target space (see Eq. \eqref{gb}).
 
	The coordinate part of the first term  in Eq.  \eqref{c21} is universal for the correlation functions in deformed CFT \cite{Cardy:1989da,Ginsparg:1988ui,Zamolodchikov:1989we}, while the field dependent coefficient represents the OPE data. 
	The expression in the bold squared parentheses in (\ref{c21}) are bilinear operations, which are explicitly given by 
	\begin{equation}\label{bG}
		\boldsymbol{[}b,G\boldsymbol{]}_{a\bar{a}}  = \partial_{\bar{c}} b_{c\bar{a}} \partial_a G^{c\bar{c}} + b_{c\bar{c}} \partial_a\partial_{\bar{a}} G^{c\bar{c}} + G^{c\bar{c}} \partial_c\partial_{\bar{c}} b_{a\bar{a}} + 
		\partial_c b_{a\bar{c}} \partial_{\bar{a}} G^{c\bar{c}}
	\end{equation}
	\begin{equation}\label{gG}
		\boldsymbol{[}g,G\boldsymbol{]}^{a\bar{a}}  = g^{c\bar{c}} \partial_c\partial_{\bar{c}} G^{a\bar{a}}- \partial_{\bar{c}} g^{c\bar{a}}\partial_c G^{a \bar{c}}  + (G \leftrightarrow g).
	\end{equation}
	In (\ref{bG}), (\ref{gG}) and in what follows the brackets $\boldsymbol{[\,\,]}$ typeset in bold acting on two tensors are to be understood as a tensor product of the Lie derivatives acting in holomorphic and antiholomorphic sectors. They also have a close relation to the Courant/Dorf\-man brackets \cite{Zeitlin2008}.
	
	Therefore, to remove UV divergences we can perform the following renormalization,
	\begin{equation}
		B\to B^{(1)}  = \left(b_{a\bar{a}}  - \boldsymbol{[}b,G\boldsymbol{]}_{a\bar{a}} \log \frac{\rho^2}{\epsilon^2}\right)\partial \varphi^a \bar{\partial}\bar{\varphi}^{\bar{a}} - 
		\frac{4}{\epsilon^2}b_{a\bar{a}} G^{a\bar{a}},
	\end{equation}
	\begin{equation}
		O_g \to  O^{(1)}_g  = \left(g^{a\bar{a}} -\boldsymbol{[}g,G\boldsymbol{]}^{a\bar{a}} \log \frac{\rho^2}{\epsilon^2}\right)p_a \bar{p}_{\bar{a}}.
	\end{equation}
	Here the logarithm plays the same role as in Eqs. \eqref{one}, \eqref{log} upon the replacement of the momentum cutoff by the coordinate regulator 
	$M_{uv} \to \epsilon^{-1}$ and the renormalization point  $\mu \to \rho^{-1}$. 
	In this way, the renormalized correlation function reduces to 
	\begin{multline} \label{Og1}
		\langle B^{(1)}(x)O^{(1)}_g(y)\rangle  = \frac{\log \frac{|x-y|^2}{\rho^2}}{|x-y|^4} \left(
		g^{a\bar{a}}\boldsymbol{[}b,G\boldsymbol{]}_{a\bar{a}} + b_{a\bar{a}} \boldsymbol{[}g,G\boldsymbol{]}^{a\bar{a}} 
		\right)+ \frac{G^{c\bar{c}}\partial_c\partial_{\bar{c}}(b_{a\bar{a}}g^{a\bar{a}})}{|x-y|^4} \log \frac{R^2}{|x-y|^2}
		\\[2mm]
		+	\frac{\partial_c(b_{a\bar{a}}G^{a\bar{c}}) \partial_{\bar{c}} g^{c\bar{a}}+ \partial_c g^{a\bar{c}}\partial_{\bar{c}}(b_{a\bar{a}}G^{c\bar{a}}) -g^{c\bar{c}}\partial_{\bar{c}}(b_{a\bar{a}}\partial_cG^{a\bar{a}})-g^{c\bar{c}}\partial_{c} (b_{a\bar{a}}\partial_{\bar{c}}G^{a\bar{a}})}{|x-y|^4}
		 + O(G^2) .
	\end{multline}
	Proceeding in the same way  step by step one can derive renormalization in the higher orders. 
	Unfortunately, all expressions become more and more bulky; besides, as we see, a lot of other terms are produced that are not responsible for the renormalization. 
	To avoid at least part of these problems we change our approach a bit. 
	
	Namely, since we are mostly interested in the renormalization of $O_g$, we limit integration to $|z|<R$ and assume that $|y|<R$ but $|x|\gg R$. 
	In this sense the perturbation does not ``touch'' observable $B(x)$. 
	Then in the leading order we immediately obtain the $O_g$ renormalization directly
	\begin{equation}
		\int\limits_{|z|<R} \frac{d^2z}{\pi} \Big\langle B(x)O_g(0)O_G(z)\Big\rangle_{0}  =\frac{b_{a\bar{a}}\boldsymbol{[}g,G\boldsymbol{]}^{a\bar{a}}}{|x|^4} \int\limits_{|z|<R} \frac{d^2z}{\pi}  
		\frac{1}{|z|^2}  =\frac{b_{a\bar{a}}\boldsymbol{[}g,G\boldsymbol{]}^{a\bar{a}}}{|x|^4} \log \frac{R^2}{\epsilon^2} 
	\end{equation}
	Here we have put coordinate of $O_g$ to zero $y=0$ because in the leading order of $|x|\to \infty$ the rest of the correlation function retains transnational invariance. 
	In the next order we have to compute 
	\begin{equation}
		C_2^{(2)}(x) =   \frac{1}{2} \int\limits_{\Sigma_\epsilon} \frac{d^2z_1}{\pi}  \frac{d^2z_2}{\pi} \Big\langle B(x)O_g(0)O_G(z_1)O_G(z_2)\Big\rangle_{0}  
	\end{equation}
	The integration domain $\Sigma_\epsilon$ is defined as 
	\begin{equation}\label{Sigma}
		\epsilon\le |z_1|\le R,\qquad \epsilon\le |z_2|\le R, \qquad \epsilon\le |z_1-z_2|\le R.
	\end{equation}
	Note extra IR regularization $|z_1-z_2|\le R$; it does not affect the UV divergences but significantly simplifies computations as it makes $\Sigma_\epsilon$ invariant
	under various changes of variables , for example, $z_1 \leftrightarrow z_2$; $z_1 \to \bar{z}_1$, $z_2 \to \bar{z_2}$; 
	$z_1 \to z_2 -z_1$, $z_2\to z_2$ etc. Using this invariance and routinely contracting $p$ and $\varphi$ using OPE \eqref{OPE1} and the Wick's theorem  
	we observe that only four possible structures appears in $C_2^{(2)}(x)$
	\begin{equation}
		C_2^{(2)}(x) = \frac{1}{|x|^4}\sum\limits_{i=1}^4 \mathcal{A}_i I_i
	\end{equation}
	with
	\begin{equation}\label{I12}
		I_1 = 	 \int\limits_{\Sigma_\epsilon} \frac{d^2z_1}{\pi}  \frac{d^2z_2}{\pi} \frac{1}{|z_1|^2|z_2|^2}, \qquad 
		I_2 =  \int\limits_{\Sigma_\epsilon} \frac{d^2z_1}{\pi}  \frac{d^2z_2}{\pi} \frac{1}{z_2^2\bar{z}_1(\bar{z}_2-\bar{z}_1)} 
	\end{equation}
	\begin{equation}\label{I34}
		I_3 = \int\limits_{\Sigma_\epsilon} \frac{d^2z_1}{\pi}  \frac{d^2z_2}{\pi} \frac{1}{z_1^2\bar{z}_2^2}, \qquad 
		I_4 = \int\limits_{\Sigma_\epsilon} \frac{d^2z_1}{\pi}  \frac{d^2z_2}{\pi} \frac{1}{|z_1|^4}.
	\end{equation}
		In the appendix \eqref{IntA} we demonstrate that the integrals can be expressed as 
	\begin{equation}
		I_1 = L_\epsilon^2 ,\qquad I_2 = L_\epsilon,\qquad I_3= 0,\qquad I_4 = R^2/\epsilon^2.
	\end{equation} 
	Here $$L_\epsilon\equiv \log R^2/\epsilon^2\,.$$ 
	The corresponding tensor structures $\mathcal{A}_i$ read as
	\begin{equation}
		\mathcal{A}_1 = \frac{b_{a\bar{a}}\boldsymbol{[}g,\boldsymbol{[}G,G\boldsymbol{]}\boldsymbol{]}^{a\bar{a}}}{4}+ \frac{b_{a\bar{a}}\boldsymbol{[}\boldsymbol{[}g,G\boldsymbol{]},G\boldsymbol{]}^{a\bar{a}}}{2}, \qquad \mathcal{A}_2 = \frac{1}{2}b_{a\bar{a}}\boldsymbol{[}g,G,G\boldsymbol{]}^{a\bar{a}}
	\end{equation}
	\begin{equation}
		\mathcal{A}_3 = b_{a\bar{a}} (\partial_c G^{k\bar{a}}\partial_k \partial_{\bar{k}} G^{c\bar{c}} \partial_{\bar{c}}g^{a\bar{k}}+\partial_c g^{k\bar{a}}\partial_k \partial_{\bar{k}} G^{c\bar{c}} \partial_{\bar{c}}G^{a\bar{k}}), 
  \end{equation}
  \begin{equation} \mathcal{A}_4 = \frac{b_{a\bar{a}}}{2} (g^{a\bar{a}}\partial_k\partial_{\bar{k}}G^{c\bar{c}} \partial_c\partial_{\bar{c}}G^{k\bar{k}} +2
		G^{a\bar{a}}\partial_k\partial_{\bar{k}}G^{c\bar{c}} \partial_c\partial_{\bar{c}}g^{k\bar{k}})
	\end{equation}
	where quadratic brackets are the operations defined as above \eqref{bG}, \eqref{gG} that the qubic bracket is given by
	\begin{multline}\label{gGG}
		\boldsymbol{[}g,G,G\boldsymbol{]}^{a\bar{a}} = \partial_{\bar{k}}G^{c\bar{c}} \partial_{\bar{c}} G^{k\bar{k}} \partial_c\partial_k g^{a\bar{a}}+ \partial_kG^{c\bar{c}} \partial_c G^{k\bar{k}}\partial_{\bar{c}}\partial_{\bar{k}} g^{a\bar{a}} + 
		2\partial_k G^{c\bar{c}} \partial_c g^{k\bar{k}} \partial_{\bar{k}}\partial_{\bar{c}} G^{a\bar{a}} +\\[2mm]  2 \partial_{\bar{k}}G^{c\bar{c}}\partial_{\bar{c}} g^{k\bar{k}}
		\partial_k\partial_c G^{a\bar{a}}
		+\partial_c\partial_{\bar{c}} G^{k\bar{k}} (\partial_k g^{c\bar{c}}\partial_{\bar{k}} G^{a\bar{a}}+ \partial_{\bar{k}}g^{c\bar{c}}\partial_k G^{a\bar{a}}+
		\partial_k g^{a\bar{a}}\partial_{\bar{k}} G^{c\bar{c}}+ \\[2mm] \partial_{\bar{k}}g^{a\bar{a}}\partial_k G^{c\bar{c}}
		+g^{a\bar{c}}\partial_k\partial_{\bar{k}} G^{c\bar{a}} + 
		G^{a\bar{c}}\partial_k\partial_{\bar{k}} g^{c\bar{a}} 
		+g^{c\bar{a}}\partial_k\partial_{\bar{k}} G^{a\bar{c}}
		+G^{c\bar{a}}\partial_k\partial_{\bar{k}} g^{a\bar{c}}	)
		\\[2mm]
		+\partial_c\partial_{\bar{c}} g^{k\bar{k}}(\partial_k G^{c\bar{c}}\partial_{\bar{k}} G^{a\bar{a}}+ \partial_{\bar{k}}G^{c\bar{c}}\partial_k G^{a\bar{a}}+
		G^{c\bar{a}} \partial_k\partial_{\bar{k}} G^{a\bar{c}} + 
		G^{a\bar{c}} \partial_k\partial_{\bar{k}} G^{c\bar{a}} 
		)
		\\[2mm]
		-(g^{k\bar{a}} \partial_{\bar{k}} G^{c\bar{c}}+G^{k\bar{a}} \partial_{\bar{k}} g^{c\bar{c}}) \partial_k\partial_c\partial_{\bar{c}} G^{a\bar{k}}- (g^{a\bar{k}} \partial_k G^{c\bar{c}} +G^{a\bar{k}} \partial_k g^{c\bar{c}})\partial_{\bar{k}}\partial_c\partial_{\bar{c}}  G^{k\bar{a}} 
		\\[2mm]
		-2 \partial_k\partial_{\bar{c}} G^{a\bar{k}} (\partial_{\bar{k}} g^{c\bar{c}}\partial_c G^{k\bar{a}}+\partial_{\bar{k}} G^{c\bar{c}}\partial_c g^{k\bar{a}})
		-2 \partial_c\partial_{\bar{k}} G^{k\bar{a}}(\partial_k g^{c\bar{c}} \partial_{\bar{c}}G^{a\bar{k}}+\partial_k G^{c\bar{c}} \partial_{\bar{c}}g^{a\bar{k}})
		\\[2mm]
		-G^{k\bar{a}} \partial_{\bar{k}} G^{c\bar{c}} \partial_k\partial_c\partial_{\bar{c}} g^{a\bar{k}} - G^{a\bar{k}} \partial_k G^{c\bar{c}} \partial_{\bar{k}}\partial_c\partial_{\bar{c}}  g^{k\bar{a}}\\[2mm] -2\partial_{\bar{k}} G^{c\bar{c}} \partial_c G^{k\bar{a}}\partial_k\partial_{\bar{c}}g^{a\bar{k}} - 2 \partial_{\bar{c}} G^{a\bar{k}} \partial_k G^{c\bar{c}} \partial_c \partial_{\bar{k}} g^{k\bar{a}}.
	\end{multline}
	
	Now we have to deform $O_g$ and $O_G$ by addition of some counterterms to make the correlation function finite when $\epsilon\to 0$. 
In particular, to remove contribution from $\mathcal{A}_4$ we have to shift $O_g$ by the scalar observable 
\begin{equation}
O_g\to O_g -\frac{\partial_k\partial_{\bar{k}}G^{c\bar{c}} \partial_c\partial_{\bar{c}}g^{k\bar{k}}}{\epsilon^2}.
\end{equation}
Note that the first term in $\mathcal{A}_4$ has a disconnected character and disappears after taking into account contributions from the partition function. 
The remaining part of the correlation function reads. 
	\begin{multline}\label{corrG2}
		C_2^{(1)}(x) + C_2^{(2)}(x) = \frac{b_{a\bar{a}}}{|x|^4} \left(g^{a\bar{a}} + L_\epsilon\boldsymbol{[}g,G\boldsymbol{]}^{a\bar{a}} +
  \frac{L_\epsilon^2}{4} \boldsymbol{[}g,\boldsymbol{[}G,G\boldsymbol{]}\boldsymbol{]}^{a\bar{a}}
  \right.\\ \left.+\frac{L_\epsilon^2}{2}\boldsymbol{[}\boldsymbol{[}g,G\boldsymbol{]},G\boldsymbol{]}^{a\bar{a}}+\frac{L_\epsilon\boldsymbol{[}g,G,G\boldsymbol{]}^{a\bar{a}}}{2}\right)
	\end{multline}
To understand renormalization further we have to separate deformation of $O_g$ and $O_G$.
In typical Callan-Symanzik approach that would correspond to $\beta$ and $\gamma$ functions \cite{zJJ}. 
 For this we consider the following one-point function, 
\begin{equation}\label{C_one}
	C_1(x) = \Big\langle B(x)\exp\left(\int_{|z|<R}\frac{d^2z}{\pi} G^{a\bar{a}}(\varphi,\bar{\varphi}) p_a\bar{p}_{\bar{a}}\right)\Big\rangle_{0}   
\end{equation}
Again here it is assumed that $|x| \gg R$. 
Its expansion up to the fourth order in $G^{a\bar{a}}$  can be deduced from \eqref{corrG2}
\begin{equation}
	|x|^4C_1(x) = b_{a\bar{a}} \left(
	G + \frac{L_\epsilon}{2}\boldsymbol{[}G,G\boldsymbol{]} + \frac{L_\epsilon^2}{4} \boldsymbol{[}G,\boldsymbol{[}G,G\boldsymbol{]}\boldsymbol{]} + 
	\frac{L_\epsilon}{3!} \boldsymbol{[}G,G,G\boldsymbol{]}	
	\right)^{a\bar{a}}
\end{equation}
To make it finite as $\epsilon\to 0$ we have to deform $G$ in the following way
\begin{equation}\label{Gdef}
	G \to \tilde{G}= G - \frac{\boldsymbol{[}G,G\boldsymbol{]}}{2}\ell +   \frac{(\ell)^2}{4} \boldsymbol{[}G,\boldsymbol{[}G,G\boldsymbol{]}\boldsymbol{]} - 
	\frac{\ell}{3!} \boldsymbol{[}G,G,G\boldsymbol{]}	+ O(G^4)
\end{equation}
with $$\ell  = \log \frac{\rho^2}{\epsilon^2}\,.$$ Then 
\begin{multline}
    \tilde{G} + \frac{L_\epsilon}{2}\boldsymbol{[}\tilde{G},\tilde{G}\boldsymbol{]} + \frac{L_\epsilon^2}{4} \boldsymbol{[}\tilde{G},\boldsymbol{[}\tilde{G},\tilde{G}\boldsymbol{]}\boldsymbol{]} + 
\frac{L_\epsilon}{3!} \boldsymbol{[}\tilde{G},\tilde{G},\tilde{G}\boldsymbol{]} =\\ G - \frac{\boldsymbol{[}G,G\boldsymbol{]}}{2}L_\rho  +   \frac{(L_\rho)^2}{4} \boldsymbol{[}G,\boldsymbol{[}G,G\boldsymbol{]}\boldsymbol{]} - 
\frac{L_\rho}{3!} \boldsymbol{[}G,G,G\boldsymbol{]}	+ O(G^4),
\end{multline}
where now $$L_\rho = \log R^2/\rho^2\,.$$ Now we can study the flow of the deformed observable 
\begin{equation}\label{flow}
\frac{\partial \tilde{G}}{\partial \ell} = -  \frac{\boldsymbol{[}G,G\boldsymbol{]}}{2} +   \frac{\ell}{2} \boldsymbol{[}G,\boldsymbol{[}G,G\boldsymbol{]}\boldsymbol{]} - 
\frac{1}{3!} \boldsymbol{[}G,G,G\boldsymbol{]}	+ O(G^4) = -  \frac{\boldsymbol{[}\tilde{G},\tilde{G}\boldsymbol{]}}{2} - 
\frac{\boldsymbol{[}\tilde{G},\tilde{G},\tilde{G}\boldsymbol{]}}{3!}  + O(\tilde{G}^4) 
\end{equation}
We see that contrary to the deformation \eqref{Gdef} the flow  \eqref{flow} does not contain logarithms but only the observables. 
Following the traditional notation we call this flow a beta function
\begin{equation}
	- \frac{\partial \tilde{G}}{\partial \ell} = \beta_2(\tilde{G}) + \beta_3(\tilde{G}) + O(\tilde{G}^4)
\end{equation}
with 
	\begin{equation}\label{beta2}
\boxed{	\beta_{2}^{a\bar{a}}(G) = \frac{\boldsymbol{[}G,G\boldsymbol{]}^{a\bar{a}}}{2} = G^{c\bar{c}} \partial_c\partial_{\bar{c}} G^{a\bar{a}}- \partial_c G^{a \bar{c}} \partial_{\bar{c}} G^{c\bar{a}} }
\end{equation}

\begin{empheq}[box=\fbox]{align}
\nonumber
\beta_3^{a\bar{a}} &= \frac{\boldsymbol{[}G,G,G\boldsymbol{]}^{a\bar{a}}}{3!} =
\frac{1}{2}\left(\partial_{\bar{k}} G^{c\bar{c}} \partial_c (\partial_k G^{a\bar{a}} \partial_{\bar{c}}G^{k\bar{k}} - G^{k\bar{a}}\partial_k \partial_{\bar{c}} G^{a\bar{k}})\right.
 \\ \nonumber
+& \partial_k G^{c\bar{c}} \partial_{\bar{c}} ( \partial_{\bar{k}} G^{a\bar{a}}\partial_c G^{k\bar{k}}-G^{a\bar{k}}\partial_c\partial_{\bar{k}} G^{k\bar{a}})
+ \\  \label{beta3}
&\left. \partial_k\partial_{\bar{k}} G^{a\bar{c}}  (G^{c\bar{a}}\partial_c\partial_{\bar{c}}G^{k\bar{k}} - \partial_{\bar{c}}G^{c\bar{k}}\partial_c G^{k\bar{a}})
+
\partial_k\partial_{\bar{k}} G^{c\bar{a}}(
G^{a\bar{c}} \partial_c\partial_{\bar{c}} G^{k\bar{k}} -\partial_c G^{k\bar{c}} \partial_{\bar{c}} G^{a\bar{k}})\right)
\end{empheq}
Once we have figured out renormalization of $G^{a\bar{a}}$ we can determine how to renormalize  $g^{a\bar{a}}$ using Eq. \eqref{corrG2}
\begin{equation}
	g \to \tilde{g} = g - \ell \boldsymbol{[}g,G\boldsymbol{]}+ \frac{\ell^2\boldsymbol{[}\boldsymbol{[}g,G\boldsymbol{]} ,G\boldsymbol{]}}{2} +\frac{\ell^2\boldsymbol{[}g,\boldsymbol{[}G,G\boldsymbol{]}\boldsymbol{]}}{4} - \frac{\ell}{2} \boldsymbol{[}g,G,G\boldsymbol{]} + O(G^3) .
\end{equation}
The corresponding flow is 
\begin{equation}
	-\frac{\partial g}{\partial \ell } = \boldsymbol{[}g,G\boldsymbol{]} + \frac{1}{2}\boldsymbol{[}g,G,G\boldsymbol{]} + O(G^3).
\end{equation}

\subsection{Results and discussion}
\label{rad}

Now let us discuss our results \eqref{beta2} and \eqref{beta3}. It was shown already in \cite{Losev2006} that  the formula \eqref{beta2}  leads to results identical to the one-loop expressions obtained in the background field method 
\cite{Callan1985,Callan1986,Fradkin1985}. 
For a K\"ahler metric and in the absence of a dilaton the two-loop $\beta$ function reduces to the  simple geometric flow that was already mentioned in the introduction \cite{Friedan1980,Friedan1985}
	\begin{equation}\label{flow2}
	\mu \frac{\partial g_{ij}}{\partial \mu} = \beta^{(1)}_{ij} + \beta^{(2)}_{ij}+ \dots 
\end{equation}
with
\begin{equation}
	\beta^{(1)}_{ij} = R_{ij},\qquad 	\beta^{(2)}_{ij} = \frac{1}{2} R_{iklm}R_{j}\,^{klm}\equiv S_{ij}.
	\label{245}
\end{equation}
Here the superscripts indicate the number of loops, while the subscripts in \eqref{beta2}, \eqref{beta3} reflect the degree of $G$. 

For a K\"ahler metric  one can easily demonstrate that $R^{a\bar{a}}=\beta_2^{a\bar{a}}$ (in Eq. \eqref{beta2}). Note that contrary to \cite{Losev2006}
we do not require neither a dilaton or the transversality condition \eqref{Trans}. 

To compare the second loop answer \eqref{beta3} we need to find $S^{ij}= G^{ik}S_{kl}G^{lj}$, which for the generic K\"ahler metrics is
\begin{multline}\label{S}
	S^{a\bar{a}} = \frac{1}{2}G_{s\bar{s}} 
	\left[
	G^{k\bar{k}}G^{c\bar{c}} \partial_c\partial_{\bar{k}} G^{a\bar{s}} \partial_k\partial_{\bar{c}} G^{s\bar{a}} 
	- G^{c\bar{c}} \partial_{\bar{k}} G^{k\bar{s}} \partial_c G^{a\bar{k}} \partial_k\partial_{\bar{c}} G^{s\bar{a}} \right.\\ \left.
	- G^{c\bar{c}} \partial_k G^{s\bar{k}} \partial_{\bar{c}} G^{k\bar{a}} \partial_c \partial_{\bar{k}} G^{a\bar{s}} 
	+ \partial_c G^{a\bar{k}} \partial_{\bar{c}} G^{c\bar{a}} \partial_k G^{s\bar{c}} \partial_{\bar{k}} G^{k\bar{s}}
	\right].
\end{multline}

This expression \textit{does not } coincide with \eqref{beta3}. This fact is immediately obvious for the one-dimensional target space, when $G^{1\bar{1}}\equiv G$. 
In this case  \eqref{beta2} and \eqref{beta3} read 
	\begin{equation}
	\beta_{2} = G \partial\bar{\partial}G - \partial  G\bar{\partial}G
\end{equation}
\begin{equation}
	2\beta_3 = (\bar{\partial}G)^2\partial^2 G + (\partial G)^2 \bar{\partial}^2 G - G \bar{\partial}G \partial^2\bar{\partial}G - G \partial G \bar{\partial}^2\partial G 
	-2 \partial\bar{\partial}G \partial G \bar{\partial}G + 2 G (\partial \bar\partial G)^2
\end{equation}
They are obviously, (differential) polynomials in $G$. The geometric answer, however, reads as 
\begin{equation}
	R^{1\bar{1}} = G\partial\bar\partial G - \partial G\bar{\partial}G,
 \end{equation}
 \begin{equation}
	 S^{1\bar{1}} = \frac{(G\partial\bar\partial G - \partial G\bar{\partial}G)^2}{G} = G(\partial\bar\partial G)^2-2 \partial G\bar{\partial}G\partial\bar\partial G  +  \frac{(\partial G\bar{\partial}G)^2}{G} 
\end{equation}
which is not even polynomial because of the last term in $S^{1\bar{1}}$. 

It is interesting to note that for $\mathds{C}P^1$ model, which corresponds to $G = \mathcal{R} (1+\varphi\bar\varphi)^2/2$, both approaches give the same answer
	\begin{equation}
	\beta_{2} = \mathcal{R}^2(1+\varphi\bar\varphi)^2/2,\qquad \beta_{3} = \mathcal{R}^3(1+\varphi\bar\varphi)^2/2.
\end{equation}
Here parameter $\mathcal{R}$ is a scalar curvature, which is constant for  $\mathds{C}P^1$, the RG flow basically 
expresses how $\mathcal{R}$ depends on $\ell$. 
In the general case, it is no longer true, and beta functions in different approaches start to be different. 
More specifically, if we focus on a slightly deformed case (see Sec. \ref{sec:currents}).
\begin{equation}\label{G3}
G = n_1 + n_2\varphi\bar{\varphi} + n_3(\varphi\bar{\varphi})^2,
\end{equation}
Then geometric approach gives 
\begin{equation}
    -\frac{\partial G}{\partial \ell} = G \left(
    \frac{\mathcal{R}}{2}+\frac{\mathcal{R}^2}{4}+ \dots 
    \right).
    \label{253}
\end{equation}
Here 
$$\frac{\mathcal{R}}{2 }= \frac{n_1n_2 +4n_1n_3 \varphi\bar{\varphi}+n_2n_3(\varphi\bar{\varphi})^2}{n_1 + n_2\varphi\bar{\varphi} + n_3(\varphi\bar{\varphi})^2}$$ 
is a non-constant curvature for the metric with the lower indices (i.e. $1/G$, with $G$ given in (\ref{G3})).
One can even conjecture that in the appropriate scheme the flow can be resummed and in this way is defined only by a one-loop result 
\begin{equation}\label{flow22}
     -\frac{\partial G}{\partial \ell} = \frac{G \mathcal{R}}{2 -\mathcal{R}}
\end{equation}
In any case, we see that the structure \eqref{253} is \textit{not} reproduced by Eqs. \eqref{beta2} and  \eqref{beta3} beyond one loop. On the contrary Eqs. \eqref{beta2}, \eqref{beta3}
show that the model is renormalizable and the corresponding flow reads
	\begin{equation}\label{flow3}
	-\frac{ \partial n_1}{\partial \ell} = n_1n_2 +  n_1 n_2^2 ,\qquad 
	-\frac{ \partial n_2}{\partial \ell} = 4 n_1n_3  +   4  n_1 n_2 n_3 ,\qquad 
	-\frac{ \partial n_3}{\partial \ell} = n_2n_3 +   n_3 (8 n_1 n_3 -n_2^2 ).
\end{equation}
This flow is interesting by itself as the stable point there depends on initial conditions. 

For two-dimensional complex target-space we see that similar to $\mathds{C}P^1$ case of Fubini-Study (Bergman) metrics Eq. \eqref{S} and \eqref{beta3} gives identical results. 
However, already for Eguchi-Hanson metric\footnote{We use notations of \cite{lye2023detailed}.}, which reads explicitly as 
\begin{equation} \label{EH}
	G_{a\bar{c}} = \sqrt{1+\frac{\alpha}{u^2}} \left(\delta_{a\bar{c}} - \frac{\alpha}{\alpha + u^2} \frac{\bar{z}_{a}z_c}{u}\right),\qquad 
	u = |z_1|^2 + |z_2|^2,
\end{equation}
one can see the difference between \eqref{flow3} and \eqref{flow22}.  Notice that  metric \eqref{EH} is a K\"ahler metric, it has unit determinant and obeys transversality condition 	$\partial_a G^{a\bar{a}} = \partial_{\bar{a}} G^{a\bar{a}} = 0$,
which immediately rules out ambiguities with a dilaton and Kalb-Ramond field.  

Finally, let us discuss the transformational properties of the obtained answers \eqref{beta2}, \eqref{beta3}.  
These expressions are manifestly not symmetric with respect to \textit{classical} diffeomorphisms of the target. 
For example the change $\varphi \to 1/\varphi$ and $\bar{\varphi} \to 1/\bar{\varphi}$ would result in the exchange $n_1 \leftrightarrow n_3$ in \eqref{G3}
which obviously is not a symmetry of Eq. \eqref{flow3} (but it is for the flow \eqref{flow22}). 
Alternatively, $n_1$ can be brought to the same value as $n_3$ by rescaling $\varphi \to \lambda \varphi$, which is also not reflected 
in \eqref{flow3}.

To quantify general case a bit more we focus on the holomorphic transformations 
\begin{equation}\label{diff}
	\varphi^a \to \varphi^a  - V^{a}(\varphi),\qquad \bar\varphi^{\bar{a}} \to \bar\varphi^{\bar{a}}. 
\end{equation}
The classical diffieormorphisms 
\begin{equation}
	G^{a\bar{a}} \to G^{a\bar{a}} + V^k \partial_k G^{a\bar{a}} - \partial_kV^a G^{k\bar{a}} 
\end{equation}
preserves the one-loop result \eqref{beta2}. Namely, the transformation properties of \eqref{beta2} are the same as for the metric
\begin{equation}
	\beta_2^{a\bar{a}} \to \beta_2^{a\bar{a}} + V^k \partial_k \beta_2^{a\bar{a}} - \partial_kV^a \beta_2^{k\bar{a}}.
\end{equation}
while $\beta_3$ in Eq. \eqref{beta3} transforms as
\begin{equation}\label{beta3T}
	\beta_3^{a\bar{a}} \to \beta_3^{a\bar{a}} + V^k \partial_k \beta_3^{a\bar{a}} - \partial_kV^a \beta_3^{k\bar{a}} -\frac{1}{2} \partial_{m}\partial_c V^{k} L_k^{mc;a\bar{a}} 
\end{equation} 
where 
\begin{multline}\label{LL}
	L_k^{mc;a\bar{a}}  =
\partial_k\partial_{\bar{k}} G^{c\bar{a}} (G^{a\bar{c}}\partial_{\bar{c}}G^{m\bar{k}}- G^{m\bar{c}}\partial_{\bar{c}} G^{a\bar{k}}) 
 + \partial_{\bar{k}} G^{m\bar{a}} (G^{a\bar{c}} \partial_k\partial_{\bar{c}}G^{cd\bar{k}} - \partial_k G^{c\bar{c}} \partial_{\bar{c}} G^{a\bar{k}})
 \\
	+\partial_kG^{m\bar{k}}\partial_{\bar{k}} ( G^{c\bar{c}}\partial_{\bar{c}} G^{a\bar{a}} - G^{a\bar{c}}\partial_{\bar{c}} G^{c\bar{a}} )+ 
	G^{m\bar{k}} \partial_{\bar{k}} (  \partial_{\bar{c}}G^{a\bar{a}} \partial_k G^{c\bar{c}} - G^{a\bar{c}} \partial_k\partial_{\bar{c}}G^{c\bar{a}}),
\end{multline}
which is another manifestation that \eqref{beta3} cannot be expressed via geometric structures (metric and curvature). 
Nevertheless, in the next section we discuss that for the $G^{a\bar{a}}$ that corresponds to current-current deformations satisfying the 2D current algebras, both 
\eqref{beta2}, \eqref{beta3}, and \eqref{LL} can be neatly expressed via the algebraic data.

Finally, let us note that usually the  ambiguities related to different choices of the regularization/renormalization scheme correspond to 
changes  \cite{Hull:1986hn}
\begin{equation}
	\beta_{\mu\nu} \sim \beta_{\mu\nu} + \nabla_\mu \xi_\nu + \nabla_\nu\xi_\mu.
\end{equation}
In the problem under consideration, it is impossible to choose a vector $\xi_\mu$ to cancel non-tensor structures, i.e. those that do not transform as tensors under the diffeormophisms. 
Moreover, it is even expected that at the quantum level, the diffeomorphisms are deformed \cite{BLN}.
The full clarification of this issue requires a separate investigation.

\section{Lie-Algebraic Sigma models}\label{sec:currents}

One of the important examples of the perturbed first-order sigma models comes from the parametrization of the $G^{a\bar{a}}$ with holomorphic and antiholomorphic vector fields, namely 
\begin{equation}
	G^{a\bar{a}} = G^{A\bar{A}} V^a_{A} V^{\bar{a}}_{\bar{A}},
\end{equation}
or equivalently $O_G =  G^{A\bar{A}} J_A \bar{J}_{\bar{A}}$, which is a current-current deformation with currents given by  $J_A = p_aV_A^a$ (and similarly for antiholomorphic). For a linear realization of the vector fields we are in a position of the so-called bosonic Gross-Neveu model \cite{Affleck2022,Bykov2021,bykov2022betafunction}. 
The current algebra immediately follows from the OPE 
\begin{equation}\label{jj1}
	J_A(z)J_B(0) = -\frac{\eta_{AB}}{z^2}+ \frac{f^C_{AB}J_C(0)- \Omega_{AB}}{z} + {\rm reg} 
\end{equation}
and similarly for the antiholomorphic components. Here we assume that vector fields form an algebra with the structure constants $f_{AB}^C$, i.e. $V^c_{[A} \partial_c V^a_{B]} \equiv  f^C_{AB} V_C^a$,
the rest of the structures are defined as 
\begin{equation}\label{etaAB}
	\eta_{AB} = \partial_k V^c_A \partial_c V_B^k,\qquad 
	\Omega_{AB} = \partial_c V_B^k\partial_m\partial_k V^c_A  \partial \varphi^m.
\end{equation}	
With such expressions the $\beta$ functions \eqref{beta2} and \eqref{beta3} reads 
\begin{equation}\label{beta22}
	\boxed{\beta_2 = \frac{1}{2} G^{A\bar{A}}G^{B\bar{B}} f^C_{AB}f^{\bar{C}}_{\bar{A}\bar{B}}}
\end{equation}
\begin{equation}\label{beta33}
	\boxed{\beta_3 = \frac{1}{2}G^{A\bar{A}}G^{B\bar{B}}G^{C\bar{C}} \left( f_{AC}^Df_{BD}^E\eta_{\bar{A}\bar{B}}  + {\rm c.c.}	\right)  }
\end{equation}
More specifically, to get \eqref{beta2} (\eqref{beta3})  from \eqref{beta22} (\eqref{beta33}) one has to contract with the corresponding vector fields $\beta^{a\bar{a}} = \beta_{C\bar{C}}V_C^aV_{\bar{C}}^{\bar{a}}$. 
These expressions correctly reproduce answers of known current-current deformations 	\cite{PhysRevLett.86.4753,Ludwig2003} (see Appendix \eqref{current}). 
They were also obtained in \cite{Sfetsos2014} using techniques developed in \cite{Sfetsos2014a} for some special choices of current algebras (see also \cite{Sagkrioti2018}).
Moreover the expression \eqref{gGG} also simplifies significantly 
\begin{equation}
    \boldsymbol{[}g,G,G\boldsymbol{]}^{a\bar{a}} = \left(G^{A\bar{A}}G^{B\bar{B}}g^{C\bar{C}}+G^{A\bar{A}}g^{B\bar{B}}G^{C\bar{C}}+g^{A\bar{A}}G^{B\bar{B}}G^{C\bar{C}}\right)\eta_{\bar{A}\bar{B}}f_{AC}^Df^E_{BD}V_E^aV_{\bar{C}}^{\bar{a}} +{\rm c.c.}
\end{equation}

Much more important is that formulas \eqref{beta22} and \eqref{beta33} follow immediately from the operator algebra and are universal. 
To be more precise  and introduce the rest of the field in \eqref{deform} we assume that fields $V_A$ ($V_{\bar A}$) form a basis in all operator fields 
and introduce a conjugate basis of 1-forms normalized such that 
\begin{equation}
	\omega_a^AV^a_B = \delta^A_B.
\end{equation}
This is equivalent for a target space to be a complex Lie group. 
The basis condition can be relaxed by considering only a certain subset of all vector fields. In this case we limit ourselves to consideration of very specific perturbations.

Provided with the basis of one-forms we introduce additional currents  
\begin{equation}
	I^A = \omega^A_a \partial \varphi^a,\qquad 	\bar{I}^{\bar{A}} = \omega^{\bar{A}}_{\bar{a}} \bar{\partial} \varphi^{\bar{a}}.
\end{equation}
This way the rest of the fields in \eqref{deform} can be presented as 
\begin{equation}
	O_\mu=m^A_{\bar{A}} J_A \bar{I}^{\bar{A}}
	,\qquad 	O_{\bar{\mu}} =\bar{m}_A^{\bar{A}} \bar{J}_{\bar{A}} I^A,\qquad O_b = b_{A\bar{A}}I^A\bar{I}^{\bar{A}}.
\end{equation}
Additionally to \eqref{jj1} the current algebra reads
\begin{equation}
	J_A(z) I^B(0) = \frac{\delta_{A}^B}{z^2} + \frac{f_{CA}^BI^C(0)}{z}+ {\rm reg},
\end{equation}
\begin{equation}
	I^B(z) J_A(0)  =\frac{\delta_{A}^B}{z^2} - \frac{f_{CA}^BI^C(0)}{z}+ {\rm reg},
\end{equation}
\begin{equation}
	I^B(z)I^C(0) = {\rm reg},
\end{equation}
and similarly in the antiholomorphic sector. 

To calculate the corresponding $\beta$ functions in the same coordinate approach as  in the previous section one has to consider in particular the correlation function 
\begin{equation}\label{c1}
	C_1(x) =\Big\langle  I^C(x) \bar{I}^{\bar{C}}(x) \exp\left(\int_{|z|<R}\frac{d^2z}{\pi}G^{A\bar{A}} J_A J_{\bar{A}}(z)\right)\Big\rangle_0 
\end{equation}
instead of \eqref{C_one}. 
The current correlation functions readily follow from the operator algebra. 
Namely, the two-point correlation function are just the $1/z^2$ coefficients, 
\begin{equation}
	\langle J_A(z)J_B(0) \rangle_0  = -\frac{\eta_{AB}}{z^2},\qquad \langle J_A(z) I^B(0)\rangle_0 = \frac{\delta_{A}^B}{z^2},\qquad \langle I^B(z)I^C(0)\rangle_0 = 0. 
\end{equation}

Higher order correlation functions can be computed using a typical trick with the contour integrations (see Appendix \eqref{JJJ}). 
In particular, three and four point correlation functions read 
\begin{equation}
	\langle I^C(x) J_A(z)J_B(w) \rangle_0 = 
	\frac{f_{AB}^C}{(x-z)(x-w)(z-w)}
\end{equation}
\begin{equation}\label{J4}
	\langle I^C(x) J_{A_1}(z_1)J_{A_2}(z_2)J_{A_3}(z_3) \rangle_0 =
	- \frac{\delta^{C}_{A_1}\eta_{A_2A_3}}{(x-z_1)^2z_{23}^2}  + \frac{f_{DA_1}^Cf_{A_2A_3}^D}{(x-z_1)z_{12}z_{13}z_{23}} + {\rm c.p.}
\end{equation}
where $ {\rm c.p.}$ stands for cyclic permutation of $1, 2, 3$. Note that since 
\begin{equation}
f_{DA_1}^Cf_{A_2A_3}^D + {\rm c.p.} =0
\end{equation}
this correlation function is $O(1/x^2)$ as should be. 
After this, the computation goes identically to the previous section and one can ``independently'' recover the results \eqref{beta22} and \eqref{beta33}.    

Now let us comment on the universal nature of the obtained results. 
In particular, we can immediately get beta function of the model deformed by $ O_{\bar{\mu}}  + O_{\mu}$ operators. 
This deformation leads to the flow of  $O_b$ operator \cite{Gamayun2009a,Gamayun2009b}. 
The exact beta function computed within the background field methods reads
\begin{equation}\label{betab}
	\partial_\ell b_{a\bar{a}} =  \bar{N}_{ak}^{\bar{c}}N^c_{\bar{a}\bar{k}} M_c^k \bar{M}^{\bar{k}}_{\bar{c}}
\end{equation}
where 
\begin{equation}
	M_c^k = [\delta_c^k- \bar{\mu}^{\bar{s}}_c\mu_{\bar{s}}^k]^{-1},\qquad 
	\bar{M}_{\bar{c}}^{\bar{k}} = [\delta_{\bar{c}}^{\bar{k}}- \mu^{s}_{\bar{c}}\bar{\mu}_{s}^{\bar{k}}]^{-1},
 \end{equation}
 \begin{equation}
	N^c_{\bar{a}\bar{k}} = \partial_{[\bar{a}} \mu^c_{\bar{k}]} - \mu^s_{[\bar{a}} \partial_s \mu^c_{\bar{k}]},\qquad 
	\bar{N}^{\bar{c}}_{ak} = \partial_{[a} \bar{\mu}^{\bar{c}}_{k]} - \bar{\mu}^{\bar{s}}_{[a} \partial_{\bar{s}} \bar{\mu}^{\bar{c}}_{k]}.
\end{equation}
Here the brackets means antisymmetrization, for instance $
	\partial_{[\bar{a}} \mu^c_{\bar{k}]}  = \partial_{\bar{a}} \mu^c_{\bar{k}} - \partial_{\bar{k}} \mu^c_{\bar{a}}
$. Notice that the beta function \eqref{betab} is in fact identical to Eq. (5.8) in \cite{Sfetsos2014}  upon identification $m \to \lambda$ and $\bar{m} \to \lambda^T$. Contrary to those findings, here Eq.  \eqref{betab} is exact since in the background field method only one-loop diagrams are non-zero \cite{Gamayun2009a}, similar to other ``all-loop'' beta functions \cite{Kutasov1989,PhysRevLett.86.4753,Itsios:2014lca,Georgiou:2016iom}.
On the other hand, the computation of the coordinate beta-function reveals a similar disagreement between the geometric answer \eqref{S} and \eqref{beta3} \cite{Gamayun2009b}.
In  \cite{Gamayun2009b} this was attributed to the seemingly different ``IR'' reason for the logarithms to appear. 
Here we can highlight the differences within in algebraic data. 
Namely, the exact beta function expanded to the second order gives 
\begin{equation}
	\beta = 	\beta_{a\bar{a}}\partial \varphi^a \bar{\partial}\bar{\varphi}^{\bar{a}} = \left(
	m^{A}_{\bar{A}}\bar{m}^{\bar{B}}_B f_{AC}^B f_{\bar{B}\bar{C}}^{\bar{A}}
	+
	f_{AB}^E f^D_{EC} (m\bar{m})^{B}_D m^A_{\bar{C}}
	+\dots\right)	I^C\bar{I}^{\bar{C}}.
\end{equation}
Repeating computation in the coordinate approach for the correlation function 
\begin{equation}
	\tilde{C}_1(x) =\Big\langle  J_C(x) \bar{J}_{\bar{C}}(x) \exp\left(\int_{|z|<R}\frac{d^2z}{\pi} (O_\mu(z)+O_{\bar{\mu}}(z))\right)\Big\rangle_0
\end{equation}
or using universal answer \eqref{beta22}, \eqref{beta33} we obtain 
\begin{equation}
	\beta_b =\left( m^{A}_{\bar{A}}\bar{m}^{\bar{B}}_B f^{B}_{AC}f^{\bar{A}}_{\bar{B}\bar{C}}+ \left(f_{AB}^E f^D_{EC} + \frac{1}{2} f_{EB}^D f_{CA}^E \right)(m\bar{m})^{B}_D m^A_{\bar{C}}+\dots \right)I^C\bar{I}^{\bar{C}}
\end{equation}
We see that in the cubic order two expressions differ by a term $\frac{1}{2} f_{EB}^D f_{CA}^E (m\bar{m})^{B}_D m^A_{\bar{C}}$. In Ref. \cite{Gamayun2009b} it was argued that the logarithms appearing in this term 
have a different nature (IR) than in front of the other terms that are responsible for the genuine UV divergences.  In the next section we exhibit that similar phenomenon occurs in the $\beta$ functions in supersymmetric theories. 

Finally, let us comment on what  the transformation law \eqref{beta3T} means in the Lie-algebraic approach. 
Namely, the terms with the second derivative of the vector field come from the transformation of $\eta_{AB}$. 
More specifically, for the generic transformation \eqref{diff} its transformation can be found from the explicit form \eqref{etaAB} 
\begin{equation}\label{deltaV}
\delta_V \eta_{AB} =  V^k \partial_k  \eta_{AB} - \partial_a \partial_k V^c V_B^k \partial_c V_A^a - \partial_c\partial_kV^a V_A^k \partial_a V_B^c 
\end{equation}
which states basically the same as transformation \eqref{beta3T}. 
We can rewrite \eqref{deltaV} equivalently as
\begin{equation}
    \delta_V \eta_{AB} = \partial_c V_A^a\partial_a (V^k \partial_k V_B^c - V_B^k\partial_k V^c) + (B \leftrightarrow A)
\end{equation}
This presentation is especially useful if $V = V_C$ is one of the fields forming the algebra. 
In this case
\begin{equation}
\delta_{V_C} \eta_{AB} = f_{CA}^D \eta_{DB} + f_{CB}^D \eta_{DA},
\end{equation}
which looks more natural than the transformation \eqref{beta3T}.

Concluding this section, let us return to subsection \ref{rad} and explain the emergence of the Lie-algebariac metric 
(\ref{G3}) from 
$sl (2)\times sl (2)$ algebra.
The   non-linear realization of $sl(2)$  generators has the form
\beq
 \ell_1=-\phi^2\,\frac{\partial}{\partial\phi}\,,\quad \ell_0= \phi\,\frac{\partial}{\partial\phi}\,,\quad \ell_{-1}= \frac{\partial}{\partial\phi}\,\nonumber\\[2mm]
\label{two}
\eeq
plus a similar equation for the  antiholomorphic algebra.

Lie-algebraic models based on $sl(2)$ are obtained as a linear
combinations of the operators
\beq
l_{-1}\bar l_{-1},\,  l_0\bar l_0,\, l_{1}\bar l_{1},\, (l_{0}\bar l_{-1} +{\rm H.c.}),\,  (l_{0}\bar l_{1} +{\rm H.c.}),\, l_{1}\bar l_{1} \
\label{three}
\eeq
with arbitrary coefficients. More precisely, 
\beq
G^{1\bar 1} \,  \frac{\partial}{\partial\phi} \frac{\partial}{\partial\bar \phi} \leftrightarrow  \sum_{a\bar b=1,-1,0} \left({\mathcal P}_{a\bar b}\, l_a \bar{l}_{\bar b}
+{\rm H.c} \right)
\label{four}
\eeq
with a set of numeric coefficients $\{ {\mathcal P}_{a\bar b} \}$.
Linear combinations of (\ref{three}) span the set of Lie-algebraic metrics $G^{1\bar 1}$. If we want to preserve a U(1) symmetry we choose the combination
\begin{equation}
n_1 l_{-1}\bar l_{-1}+ n_2 l_0\bar l_0, + n_3 l_{1}\bar l_{1}
\label{328}
\end{equation}
which is equivalent to the metric (\ref{G3}).

	\section{Comparison with Supersymmetric sigma models}
	\label{hyp}
	
	One possible way out reconciling the first-order perturbation theory with the standard geometric formalism can be presented as follows.
	The OPE in the coordinate space in the Lie-algebraic (bosonic Gross-Neveu) problem which represents
	the genuine {\em Wilsonean } OPE in one loop.
	Higher loops are obtained as a result of  certain  ``anomaly" in the measure 
	revealing itself in the calculation of the second and higher loops, in analogy with Yang-Mills theory.\footnote{Also, 
	the phenomenon we observe has a direct parallel in the chiral ${\mathcal N}=(0,2)$ two-dimensional $\mathds{C}P^1$ model \cite{CS}.}  A general proof is presented  in \cite{NSZV,NSZV1,CS,AH} while the direct perturbative all-loop calculations are given in 
	\cite{KS,KS1,KS2}. 
	
	One can complexify the coupling constants in the Lagrangians introducing their holomorphic counterpartners,
	\beqn
	&& \frac{8\pi^2}{g^2}+ i\theta\qquad {\rm super\,\,YM},\nonumber\\[2mm]
	&&\frac{4\pi}{g^2}+ i\theta\qquad {\mathcal N} =(0,2)\,\, \mathds{C}P^1.
	\label{two3}
	\eeqn
	In fact, this complexification is automatic in the superfield language. Moreover, it is maintaned in the  $\beta$-function calculations at one-loop,
	but the same supersymmetry forces it to be broken at higher loops \cite{NSZV,NSZV1,CS} due to infrared singularities for massless fields. 
	
	In other words, the {\em bona fide}  Wilsonean OPE in $\none$ super-Yang-Mills and (0,2) $\mathds{C}P^1$ model ends at one loop and is holomorphic; the holomorphic anomaly shows up at two loops in the matrix elements of the relevant operators rather than in the OPE coefficients.
	In   \cite{AH} this phenomenon in supersymmetric theories was related to an anomaly in the corresponding path integral  measure. 
	
	The exact all-order NSVZ $\beta$ function in super-Yang-Mills \cite{NSZV} and its analog in ${{\mathcal N} =(0,2)}$ $\mathds{C}P^1$ model ensued. In the both cases the two-loop result in the $\beta$ functions fixes all higher orders through a geometrical progression. Of course, 
	since the third and higher orders are scheme dependent one must use a perturbative regularization/renormalization scheme compatible with supersymmetry.
	This was directly verified for the first time in  \cite{KS,KS1,KS2}.

	\section{Instanton analysis}\label{sec:insta}

 In the previous sections, we have established that the first-order formalism for sigma models proposed in \cite{Losev2006}
when applied to $\beta$ function calculations works perfectly in the leading order, however, leads to results incompatible with the standard geometric (background field) calculations starting from the second order. The hypothesis we have formulated to explain the discrepancy is as follows: 
the polynomiality in the second and higher loops is due to an infrared effect which in turn reflects the loss of symmetry in the measure not explicitly seen in the path integral.  In our subsequent work \cite{subsequent} we will demonstrate that this is indeed the case.  Starting from a supersymmetry-based regularization it will be proven that making superpartners' mass large we are left with a finite effect, an anomaly.  In this section, we present a parallel of exactly the same phenomenon with the loss of {\em holomorphy} beyond one loop which occurs in four-dimensional Yang-Mills theory and two-dimensional $\mathds{C}P^1$ model. The latter example is especially close to the problem we deal with in the present paper. Below we use the instanton-based analysis to reveal the parallel. (It is not the only possible demonstration, but it is the simplest).
	
	In the instanton background a part of supersymmetry is preserved. The  symmetries broken by the given instanton generate zero modes. The exact formula for the instanton measure in pure SUSY-Yang-Mills (without matter) is
	\beq
	d\mu_{{\rm SU}(2)}=\frac{1}{256\pi^{2}}e^{-8\pi^{2}/g^{2}}(\mu)^{n_b-{\textstyle\frac{n_f}{  2}}}\left(\frac{8\pi^{2}}{g^{2}}\right)^{{{\textstyle\frac{1}{ 2}}(n_b-n_f)}}
	d^{4}x_0\,d^{2}\theta_{0}\,d\rho^{2}\,d^{2}\,\bar\beta
	\label{eme}
	\eeq
	where $n_b$ and $n_f$ are the numbers of the boson and fermion zero modes, respectively, $x_0,..., \bar\beta$ are the instanton  moduli, both boson and fermion\footnote{Integration over the instanton three angular orientations in the SU(2) group is performed.}. In what follows there is no need to 
	show integration over the moduli explicitly, since we will focus on the $\beta$ function.
	For the SU(2) gauge group
	\beq
	n_b=8,\,\,\, n_f=4\,.
	\eeq
	For SU$(N)$ we have
	\beq
	n_b=4N,\,\,\, n_f=2N\,.
	\eeq
	
	Since the expression (\ref{eme}) is exact we can immediately derive the $\beta$ function for $1/g^2$ to all orders.
	The $\mu$ dependence appears in (\ref{eme}) explicitly in the pre-exponent and implicitly through $g^2$. The combination as a whole must be $\mu$ independent, therefore,
	\beq
	\beta (g^2) =\frac{\pt}{\pt L} \frac{1}{g^2} = \left(n_b-\frac{n_f}{2}\right) \frac{1}{8\pi^2}\, \frac{1}{1-{\textstyle\frac{1}{  2}}\,(n_b-n_f){\textstyle\frac{g^2}{8\pi^2}}}.
	\label{ebe}
	\eeq
	The second and higher order coefficients in the $\beta$ function
	come from the denominator in (\ref{ebe}) which, in turn, is entirely determined by the  factor $\left(\frac{8\pi^{2}}{g^{2}}\right)^{{{\textstyle\frac{1}{ 2}}(n_b-n_f)}}$ in Eq. (\ref{eme}) representing infrared (zero modes) effect. Note that $g^2$ in the denominator in fact must be understood as Re$\,g^2$. The reason is obvious: the parameter  $\theta={\rm Im} \,(1/g^2) $ does not appear in perturbation theory. However, if we ignore the denominator (i.e. limit ourselves to one loop) the coupling $1/g^2$ in the left-hand side  of (\ref{ebe}) can be viewed as holomorphic. 
	Equation (\ref{ebe}) (and (\ref{ebe4}) below) is to be compared with Eq. (\ref{flow22}).
	
	Let us consider extended supersymmetry. For $\ntwo$ we have $n_f=n_b$, hence, the denominator disappears, we have one-loop $\beta$ function and the holomorphy is preserved. For $\nfour$ we have  $n_f=2n_b$, and the $\beta$ function vanishes.
	
	Adding matter fields in $\none$ Yang-Mills theory produces two changes. First, the overall first coefficient in (\ref{ebe}) must now include the matter field contributions. 
	Generically it still remains integer and holomorphy is preserved. Second, the anomalous dimensions of the matter fields appear in the NSVZ $\beta $ function starting from the {\em second} loop; they defy holomorphy since the $Z$ factors which produce the anomalous dimensions are non-lolomorphic.
	
	We make a pause here to rewrite (\ref{eme}) in the following useful general form,
	\beq
	d\mu_{{\rm SU}(N)}=\propto e^{-S_{\rm inst} }(\mu)^{n_b-{\textstyle\frac{n_f}{  2}}}\left(S_{\rm inst}\right)^{{{\textstyle\frac{1}{ 2}}(n_b-n_f)}}
	\label{eme2}
	\eeq
	where $S_{\rm inst}$ is the instanton classical action $S=8\pi^2/g^2$.
	
	\vspace{2mm}
	
	Now, it is instructive to briefly review the ${\mathcal N}=(0,2) $ $\mathds{C}P^1$ model  since it exhibits a novel feature absent in Yang-Mills. This model can be formulated in terms of ${\mathcal N}=(0,2) $ superfieds. In its minimal version it is constructed from a single superfield with two 
	physical components -- a complex scalar field $\phi$ and a two-dimensional Weyl fermion $\psi_L$. The $\mathds{C}P^1$ target space is, of course, K\"ahlerian,
	which allows one to introduce a complexified coupling, see Eq. (\ref{two3}).
	The perturbative superfield calculation yields \cite{CS} in one and two loops
	\beq
	\beta (g^{-2})_{(0,2)} =\frac{\pt}{\pt L} \frac{1}{g^2} = \frac{1}{2\pi}\, \left(1+{\frac{g^2}{4\pi}}+...\right).
	\label{ebe1}
	\eeq
	While the one loop contribution preserves holomorphy in $1/g^2$, the second loop term in the parentheses is in fact ${\textstyle\frac{{\rm Re}\,g^2}{ 4\pi}}$ 
	which ruins holomorphy. This is due to an infrared singularity which can be seen both perturbatively and through the instantons. Unlike the Yang-Mills theory
	the fermion contribution in perturbation theory starts from the second loop.
	
	Let us start from the instanton in ${\mathcal N}=(2,2) $ $\mathds{C}P^1$. The latter is well known to produce only one-loop $\beta$ function 
	due to the K\"ahler nature of the target manifold \cite{AGF,AGF1,AGF2}. The instanton measure is
	\beq
	d\mu_{(2,2)\,\mathds{C}P^1} \propto e^{-4\pi/g^2}\left(\mu\right)^2 \,d x_0 \frac{d\rho}{\rho}\{d\alpha\, d\bar \alpha \,d\beta \,d\bar \beta\} 
	\label{two23}
	\eeq
	Fermion and boson zero modes completely cancel, both in the pre-exponent and in the exponent. There are four real boson zero modes and four fermion zero modes (the fermion moduli are in braces in Eq. (\ref{two2})). Integration over the U(1) angle is carried out.
	The zero modes give rise to 
	\beq
	\left(\frac{\mu}{g}\right)^{4}\quad ({\rm boson}), \qquad \left(\frac{\mu}{g^2}\right)^{-2} \quad ({\rm fermion}).
	\label{zm}
	\eeq
	All two-loop and higher loops in the instanton background cancel each other, see e.g. \cite{obz}, Sect. 6.4. Therefore there is no $g$ dependence in the pre-exponent.
	Imposing the condition of $\mu$ independence in the right-hand side of (\ref{two23}) we obtain  one-loop $\beta$ function as was expected.
	
	Now let us pass to ${\mathcal N}=(0,2) $ which is exactly half-way between (\ref{two23}) and non-supersymmetric $\mathds{C}P^1$ \cite{shifp}.
	In the passage we must completely discard the $\psi_R$ Weyl fermion present in $(2,2)$. As a result, if
	in the previous $(2,2)$ case we had four fermion zero modes, now we have two. Thus, the fermion zero mode factor 
	$$ \left(\frac{\mu}{g^2}\right)^{-2}_{(2,2)} \to \left(\frac{\mu}{g^2}\right)^{-1}_{(0,2)} ,$$
	implying $\left(\mu\right)^3$ in the pre-exponent. This would change the first coefficient in the $\beta$ function which cannot happen because fermion contribution emerges only at the second loop. What is forgotten?
	
	Two-loop and higher loops in the instanton background still cancel each other \cite{CS}. However, the balance between the boson and fermion {\em non-}zero modes characteristic to the $(2,2)$ model is destroyed by elimination of the $\psi_R$ Weyl fermion field from the theory. This produces
	extra $\exp(\log \mu )$ factor. Assembling all these factors together we arrive at  the $(0,2)$ instanton measure,
	\beq
	d\mu_{(0,2)\,\mathds{C}P^1 } \propto e^{-4\pi/g^2}\left(\mu\right)^2 \frac{1}{g^2}\,.
	\label{two2}
	\eeq
	resulting in the following exact$\beta$ function
	\beq
	\beta (g^{-2})_{(0,2)} \equiv \frac{\pt}{\pt L} \frac{1}{g^{-2}} = \frac{1}{2\pi}\, \frac{1}{\left(1-{\frac{g^2}{4\pi}}\right)}, 
	\label{ebe4}
	\eeq
	to be compared with Eq. (\ref{ebe1}).

	A couple of remarks are in order here. First, again, in the denominator we have in fact $\frac{{\rm Re}\,g^2}{4\pi}$ rather than $\frac{g^2}{4\pi}$.
	Second, in the $(0,2)$ model  one can introduce ``matter" fields much in the same way as in 4D super-Yang-Mills. Their impact is similar to that of matter in super-Yang-Mills \cite{CS}. Finally, a comparison of \eqref{ebe4} to \eqref{ebe1} gives a strong hint in favor of conjecture Eq. \eqref{flow22}, for the Lie-algebraic $sl(2)\times sl(2)$ structure \eqref{G3} \footnote{We remind that the third and higher order coefficients are scheme dependent. Although currently they are calculated up to four loops \cite{Grah} the renormalization schemes used do not necessarily preserve the 
		$sl(2)\times sl(2)$ algebraic structure.}. 
		
		In Section \ref{sec:currents} we introduced the simplest Lie-algebriac model. Its Lie-algebriac structure is presented in Eq. (\ref{328}). 
Above in this section we studied ${\mathcal N}=(0,2) $ supersymmetric generalization of $\mathds{C}P^1$ model, in combination with the instanton
analysis to argue that the $\beta$ function is of the NSVZ type. Like in ${\mathcal N}=1 $ super-Yang-Mills theory, $\mathds{C}P^1_{\,\,(0,2)}$ has only one coupling constant and one beta function. Now let us address a problem with two coupling constants and two coupled $\beta$ functions, with the metric $G^{1\bar 1}$ from Eq. (\ref{G3}) and ${\mathcal N}=(2,2) $ supersymmetry.  It is known that the K\"ahlerian nature of the target space in this case guarantees that only one-loop result survives in the instanton background.  Let us see whether we can find  both $\beta$ functions.

The metric of the ${\mathcal N}=(2,2) $ generalized model in Eq. (\ref{328}) contains three parameters, $n_{1,2,3}$. Rescaling the fields $\phi$ one can always arrange $n_1=n_3$ even if in the original metric $n_1\neq n_3$. The only condition is that both $n_{1,3}\neq 0$.  Then we are left with 
two parameters,\footnote{Note that Eq. (\ref{flow3}) does not allow this rescaling because it is incompatible with diffeomorphisms.} $g^2$ and $k$,
\begin{equation}
n_1 = n_1= n_3= \frac{g^2}{2}, \quad  n_2 = g^2k\,.
\end{equation}
Besides the moduli, the instanton measure now takes the following exact form,
\begin{equation}
d\mu_{(2,2)} \propto \mu^2 e^{-S_{\rm inst}}\,,\quad 
S_{\rm inst} =
{\frac{4\pi}{g^2}} \frac{{\rm arccosh} (k)}{\sqrt{k^2-1}} \,,
  \quad k \geq 1\,.
\label{frid}
\end{equation}
For $k<1$ the second expression in (\ref{frid}) must be analytically  continued.
From the above equation differentiating over $L=\log\mu$ we obtain an exact relation
\begin{equation}
2 - \frac{\partial}{\partial L} S_{\rm inst} = 0\,.
\label{frida}
\end{equation}
$S_{\rm inst}$ now depends on two constant. To derive two $\beta$ functions for two couplings we need another relation. 
At one loop we know that
\begin{equation}
g^4\left(k^2-1\right) = {\rm RG\,\, invariant}.
\label{friday}
\end{equation}
We may or may not assume that it holds to all orders. So far it is not yet establsihed.
Equation (\ref{friday}) in conjunction with (\ref{frida}) implies
\begin{equation}
 \beta_k  =\frac{\partial}{\partial L}k=\frac{g^2}{2\pi}(k^2-1)
 \label{fridayp}
\end{equation}
Applying ${\partial}/{\partial L}$ to (\ref{friday}) we arrive at
\begin{equation}
\beta_g = \frac{\partial}{\partial L}g^2 = -\frac{k}{k^2-1} g^2 \beta_k =- \frac{k g^4}{2\pi}\,
 \label{fridaypp}
\end{equation}
Equations (\ref{fridayp}) and (\ref{fridaypp}) coincide with the standard one-loop perturbative calculations.
For $k=1$ (i.e. undeformed $\mathds{C}P^1$) $\beta_k=0$ and $\beta_g$ coincides with that of $\mathds{C}P^1$ (here ${\alpha'=1/(2\pi)}$).

	\section{Conclusions}
	
	We applied the first order formalism of  loop calculations in generic K\"ahler 2D sigma models suggested in \cite{Losev2006} (for further developments see 
	\cite{Zeitlin2006,Zeitlin2008,Zeitlin2009})
	to  multiloop calculations of  $\beta$ functions, using the coordinate space graphs and operator product expansion (OPE) rules. We also considered the same formalism in Lie-algebraic sigma models, in which the OPE rules are especially transparent. 
	Starting from the bare Lagrangian we obtained  a certain pattern of loop coefficients $\beta^{(i)}$. In the one-loop coefficient we observed that the first-order formalism  and the old standard geometric method of the $\beta$-function calculations fully coincide. However, starting from the second loop the coincidence disappears at least for some metrics. While the standard geometric method respects {\em all} symmetries of the target space at the classical level,
	this is not the case with the first-order formalism even when  one cannot expect 
	such a paradox. In a bid to solve the discrepancy, we made a number of encouraging observations, but the full solution of the issue is still evasive. We conjectured that some so far undetected infrared anomalies might be responsible for the non-completeness of the first-order calculations, in the same vein as it happened in  the 
	anomaly supermultiplet in super-Yang-Mills. Our hypothesis is formulated in Eq. \eqref{flow22}.
	
	What remains to be done? It is highly desirable to directly verify (or falsify) the all-order formula (\ref{flow22}) which defies polynomiality of the first-order formalism predictions for Lie-algebraic K\"ahler target spaces. 
	Alternatively, we could undertake a step-by-step approach and further search for loopholes in the first-order formalism 
	focusing on the second loop. This problem is quite general, and so must be its solution. After all, taken separately, theoretical elements of the first-order formalism are solid. There is nothing wrong in Feynman graphs in the coordinate space for massless fields, or in OPE rules.
Taking into account dilaton's contribution \cite{Callan1985,Callan1986} might also ``fix'' the beta function \cite{Witten1991}. 
However, for the transversal  K\"ahler metrics these contributions are absent \cite{Losev2006} yet the coordinate and geometric approaches give different results. 
Besides we have reproduced beta functions for the current-current deformations 
without invoking any dilatons. Yet systematic clarification of the dilaton role is highly desirable.

Another question for the future is if some Lie-algebraic  metrics
	define an integrable model as it happens with $\mathds{C}P^1$ then one may expect to determine the spectrum of the asymptotic states and observe how the model at hand  acquires a mass gap generation. The number  of such states is still three
 at least, in a range of parameters close to $\mathds{C}P^1$, but, of course, the $O(3)$ symmetry of the spectrum must be lost as it is explicitly lost on the target space.
	
	\section*{Acknowledgments}
	
	We are grateful to Adam Bzowski, Alexey Litvinov and Dmitri Bykov for useful discussions.  
 We would like to thank Kostas Sfetsos, Kostas Siampos, and Arkady Tseytlin for their feedback, comments and suggestions.

The work of MS is supported in part by DOE grant DE-SC0011842.	

\vspace{5mm}

 	\appendix
	\centerline{\Large\bf Appendices}

	\section{Integrals}\label{IntA}
	
	In this appendix we study integrals \eqref{I12}, \eqref{I34} needed for the second-order perturbation theory. We start with $I_1(\epsilon)$ defined as
	\begin{equation}
		I_1 (\epsilon) =  \int\limits_{\Sigma_\epsilon} \frac{d^2z_1}{\pi}  \frac{d^2z_2}{\pi} \frac{1}{|z_1|^2|z_2|^2}
	\end{equation}
	where $\Sigma_\epsilon$ is defined in \eqref{Sigma}. Notice that $\epsilon$ dependence comes only in the form $R/\epsilon$. 
	
	To find $\epsilon$ dependence we compute the following derivative
	\begin{equation}
		-\epsilon \frac{\partial}{\partial \epsilon} I_1(\epsilon) = \epsilon\int\limits_{\Sigma_\epsilon} \frac{d^2z_1}{\pi}  \frac{d^2z_2}{\pi} \frac{\delta(|z_1|-\epsilon) + \delta(|z_2|-\epsilon) + \delta(|z_{12}|-\epsilon)}{|z_1|^2|z_2|^2}
	\end{equation}
	The delta functions fix absolute value of the corresponding variables, however the phase can be also fixed due to the rotational symmetry $z_i \to z_i e^{i\theta}$ inside the integral. 
	This way one of the integrals can be computed completely, and the result after additional rescaling $z\to \epsilon u$ leads to 
	\begin{equation}
		-\epsilon \frac{\partial}{\partial \epsilon} I_1(\epsilon) = 2\int\limits_{\hat\Sigma_1} \frac{d^2u}{\pi} 
		\left(
		\frac{2}{|u|^2} + \frac{1}{|u|^2|u-1|^2}
		\right)
	\end{equation}
	where $\hat\Sigma_1$ is defined as 
	\begin{equation}
		1\le |u| \le R/\epsilon,\qquad 1\le |u-1| \le R/\epsilon.
	\end{equation}
	we can explicitly present this integral as 
	\begin{equation}
		-\epsilon \frac{\partial}{\partial \epsilon} I_1(\epsilon) =4 \log \frac{R^2}{\epsilon^2}- 4 \int\limits_{\Sigma_1^*} \frac{d^2u}{\pi}  \frac{1}{|u|^2} +  2\int\limits_{\Sigma_1} \frac{d^2u}{\pi} 
		\frac{1}{|u|^2|u-1|^2} + O(\epsilon/R)
	\end{equation} 
	where we have denoted new integration domains as 
	\begin{equation}\label{sigma1}
		\Sigma_1: 	|u|\ge 1,\quad |u-1|\ge 1; \qquad  \Sigma_1^*:|u|>1,\quad |u-1|<1.
	\end{equation}
	Notice that since integrals are IR convergent we have effectively sent $R\to\infty$, which resulted in terms of order $O(\epsilon/R)$,
	which we further disregard. 
	Here we have also used 
	\begin{equation}
		\int\limits_{R>|u|, |u-1|>R} \frac{d^2u}{|u|^2} = O(1/R).
	\end{equation}
	
	Integral over $\Sigma_1$ can be computed for instance with the help of the  Stokes's theorem 
	
	Similarly we can present 
	\begin{multline}\label{u1}
		\int\limits_{ \Sigma_1} \frac{d^2u}{\pi}  \frac{1}{|u|^2|u-1|^2}   
		= 	\frac{i}{2\pi }\int\limits_{ \Sigma_1}   \frac{du \wedge d \bar{u}}{|u|^2|u-1|^2} = 	\frac{i}{2\pi }\int\limits_{ \Sigma_1}  d \left(
		\log \frac{u}{u-1} \frac{ d \bar{u}}{\bar{u}(\bar{u}-1)} \right)= \\ 	
		\frac{i}{2\pi } \int_{\varphi \in [\frac{\pi}{3},\frac{5\pi}{3}]} 
		\left[
		\log \frac{u}{u-1} \frac{d\bar{u}}{\bar{u}(1-\bar{u})}\Big|_{u= e^{i\varphi}} + \log \frac{u}{u-1} \frac{d\bar{u}}{\bar{u}(1-\bar{u})}\Big|_{u=1- e^{i\varphi}} 
		\right]
	\end{multline}
	which leads to 
	\begin{multline}
		\int\limits_{|u|>1, |u-1|>1} \frac{d^2u}{\pi}  \frac{1}{|u|^2|u-1|^2}   =  \int\limits_{\pi/3}^{5\pi/3} \frac{d\varphi}{\pi} \frac{\log(1-e^{-i
				\varphi})}{1-e^{-i
				\varphi}} = \\
		\int\limits_{\pi/3}^{5\pi/3} \frac{d\varphi}{2\pi i} (\cot \frac{\varphi}{2} +i)\left(\log\left(2\sin \frac{\varphi}{2}\right)+i\frac{\pi - \varphi}{2} \right) = 
		\int\limits_{\pi/3}^{5\pi/3} \frac{d\varphi}{\pi}\log\left(2\sin \frac{\varphi}{2}\right)
		\approx 0.64
	\end{multline}
	The integral over $\Sigma_1^*$ can be computed in a similar manner, or directly in the polar coordinates 
	\begin{equation}
		\int\limits_{\Sigma^*_1} \frac{d^2 u}{|u|^2} = 2\int\limits_{0}^{\pi /3} d\theta \int\limits_{1}^{2\cos \theta}  \frac{dr}{r} = 2\int\limits_{0}^{\pi /3} d\theta \log\left(2\cos\theta\right)
		= \int\limits_{\pi /3}^\pi d\theta \log\left(2\sin\frac{\theta}{2}\right).
	\end{equation}
	Using this result and \eqref{u1} we conclude that 
	\begin{equation}
		2\int\limits_{\Sigma_1} \frac{d^2u}{\pi} 
		\frac{1}{|u|^2|u-1|^2}- 4 \int\limits_{\Sigma_1^*} \frac{d^2u}{\pi}  \frac{1}{|u|^2} =0,
	\end{equation}
	which is equivalent to 
	\begin{equation}
		I_1(\epsilon) = \log^2 \frac{R^2}{\epsilon^2} + o(\epsilon/R).
	\end{equation}
	Next we turn into the integral 
	\begin{equation}
		I_2(\epsilon) = \int\limits_{\Sigma_\epsilon} \frac{d^2z_1}{\pi}  \frac{d^2z_2}{\pi} \frac{1}{z_2^2\bar{z}_1(\bar{z}_2-\bar{z}_1)} 
	\end{equation}
	Computing derivative in the same way as previously we obtain 
	\begin{equation}
		- \epsilon \frac{\partial I_2(\epsilon)}{\partial \epsilon} = 2 \int\limits_{\Sigma_1} \frac{d^2u}{\pi} \left(
		\frac{2}{u^2(\bar{u}-1)} + \frac{1}{\bar{u}(1-\bar{u})}
		\right)+ O(\epsilon/R).
	\end{equation}
	Here we have replaced $\hat\Sigma_1$ to $\Sigma_1$ defined in \eqref{u1} (and correction $O(\epsilon/R)$) because integrals are now IR convergent. 
	Applying Stokes's theorem we get 
	\begin{multline}
		\int\limits_{\Sigma_1} \frac{d^2u}{\pi} \left(
		\frac{2}{u^2(\bar{u}-1)} + \frac{1}{\bar{u}(1-\bar{u})}
		\right) = \int\limits_{\Sigma_1}  d \left[
		\left(
		\frac{2}{u(1-\bar{u})} + \frac{u}{\bar{u}(1-\bar{u})} 
		\right)\frac{d\bar{u}}{2\pi i}
		\right] \\= \int\limits_{\pi/3}^{5\pi/3} d\varphi \frac{3i - 5 \cot\frac{\varphi}{2}+8 \sin\varphi}{4\pi i} = 1.
	\end{multline}
	This gives the following estimate 
	\begin{equation}
		I_2(\epsilon) = \log \frac{R^2}{\epsilon^2} + o(\epsilon/R).
	\end{equation}
	The next integral to compute is 
	\begin{equation}
		I_3(\epsilon) = \int\limits_{\Sigma_\epsilon} \frac{d^2z_1}{\pi}  \frac{d^2z_2}{\pi} \frac{1}{z_1^2\bar{z}_2^2}.
	\end{equation}
	Computing $\epsilon$ derivative we obtain 
	\begin{equation}
		- \epsilon \frac{\partial I_3(\epsilon)}{\partial \epsilon} = 2 \int\limits_{\Sigma_1} \frac{d^2u}{\pi} \left(
		\frac{2}{u^2} + \frac{1}{(u-1)^2\bar{u}^2}
		\right)=  2\int\limits_{\partial\Sigma_1} \frac{d\bar{u}}{2\pi i } \left(
		\frac{2}{u} + \frac{1}{(u-1)\bar{u}^2}
		\right) =0
	\end{equation}
	Here we have applied Stokes's theorem and obtained zero after integration over the angle. Therefore we conclude 
	\begin{equation}
		I_3(\epsilon) = o (\epsilon/R).
	\end{equation}
	Finally, we can compute the following integral 
	\begin{equation}
		I_4(\epsilon) = \int\limits_{\Sigma_\epsilon} \frac{d^2z_1}{\pi}  \frac{d^2z_2}{\pi} \frac{1}{z_1^2\bar{z}_1^2}.
	\end{equation}
	We compute difference 
	\begin{equation}
		\delta I_4(\epsilon) =I_4(\epsilon) - \int\limits_{R\ge |z_1|, |z_2|\ge \epsilon} \frac{d^2z_1}{\pi}  \int \frac{d^2z_2}{\pi} \frac{1}{z_1^2\bar{z}_1^2} = 
		-  \int\limits_{|z_1|\ge \epsilon} \frac{d^2z_1}{\pi}  \int\limits_{|z_2|\ge \epsilon} \frac{d^2z_2}{\pi} \frac{\theta(\epsilon - |z_1-z_2|)}{z_1^2\bar{z}_1^2} + O(\epsilon/R)
	\end{equation}
	The obtained difference is IR convergent so we have dropped the upper limit. Now we can rescale the whole expression 
	\begin{equation}
		\delta I_4(\epsilon) =- \int\limits_{|z_1|\ge 1} \frac{d^2z_1}{\pi}  \int\limits_{|z_2|\ge 1} \frac{d^2z_2}{\pi} \frac{\theta(1 - |z_1-z_2|)}{z_1^2\bar{z}_1^2} + O(\epsilon/R)
	\end{equation}
	Therefore
	\begin{equation}
		I_4 (\epsilon) = \frac{R^2}{\epsilon^2} + O(1). 
	\end{equation}
	
	\section{Current-Current deformations} \label{current}
	
	In this appendix, we provide a comparison of the general formulas for $\beta_2$ and $\beta_3$ (Eqs. \eqref{beta22},\eqref{beta33}) with the exact all-order answers given in 
	\cite{PhysRevLett.86.4753,Ludwig2003}, for the group $SU(2)$ on the level $k$. 	
	More precisely, following 	\cite{PhysRevLett.86.4753} the current-current deformation is parametrized as  
	\begin{equation}
		\delta S  =\int  g_{\perp}\left( J^+ \bar{J}^- 
		+ J^- \bar{J}^+ \right) 
		+ g_{\parallel} J_3\bar{J}_3 \ ,
		\label{su2:int}
	\end{equation}
	with the current normalization as follows
	\begin{equation}
		J_3 (z)  J^{\pm} (0)  \sim
		\pm \frac{1}{z} J^{\pm} (0), \qquad
		J^+ (z) J^- (0) \sim
		\frac{k}{2}\frac{1}{z^2} + \frac{1}{z} J_3 (0),\qquad
		J_3 (z)  J_3 (0)  \sim \frac{k}{2}\frac{1}{z^2} \, . 
		\label{su2:curr}
	\end{equation}
	We require first two orders of the claimed exact beta functions there
	\begin{eqnarray}
		\beta_{g_\perp} = \frac{g_\perp 
			\left( g_\parallel - k g^2_\perp/4\right)}
		{\left( 1 - k^2g^2_\perp/16 \right) 
			\left( 1 + k g_\parallel/4 \right)} &\approx& g_\perp 
		g_\parallel -\frac{k}{4}g^2_\perp 
		g_\parallel -\frac{k}{4}g_\perp 
		g^2_\parallel + O(g^4),
		\nonumber\\
		\beta_{g_\parallel} = \frac{g^2_\perp 
			{\left( 1 - k g_\parallel/4\right)}^2}
		{{\left( 1 - k^2g^2_\perp/16 \right)}^2} &\approx& g^2_\perp -\frac{k}{2}g^2_\perp 
		g_\parallel+O(g^4). \label{beta:su2}
	\end{eqnarray}
	In our notations for $J_{A} = (J^-,J^+,J_3)$, we obtain
	\begin{equation}
		G^{A\bar{A}} = \left(
		\begin{array}{ccc}
			0 & g_\perp & 0 \\
			g_\perp & 0 & 0 \\
			0 & 0 & g_\parallel \\
		\end{array}
		\right),\qquad \eta_{AB} = \frac{k}{2}\left(
		\begin{array}{ccc}
			0 & 1 & 0 \\
			1 & 0 & 0 \\
			0 & 0 & 1 \\
		\end{array}
		\right)
	\end{equation}
	\begin{equation}
		f^3_{AB} = \left(
		\begin{array}{ccc}
			0 & 1 & 0 \\
			-1 & 0 & 0 \\
			0 & 0 & 0 \\
		\end{array}
		\right),\qquad f_{AB}^+ =\left(
		\begin{array}{ccc}
			0 & 0 & -1 \\
			0 & 0 & 0 \\
			1 & 0 & 0 \\
		\end{array}
		\right),\qquad 
		f^-_{AB} = \left(
		\begin{array}{ccc}
			0 & 0 & 0 \\
			0 & 0 & 1 \\
			0 & -1 & 0 \\
		\end{array}
		\right).
	\end{equation}
	This gives 
	\begin{equation}
		\beta_2 = \frac{1}{2}G^{A\bar{A}}G^{B\bar{B}} f^C_{AB}f^{\bar{C}}_{\bar{A}\bar{B}} = - \left(
		\begin{array}{ccc}
			0 & g_\perp g_\parallel  & 0 \\
			g_\perp g_\parallel  & 0 & 0 \\
			0 & 0 & g_\perp^2 \\
		\end{array}
		\right)
	\end{equation}
	\begin{equation}
		\beta_3 =\frac{1}{2}G^{A\bar{A}}G^{B\bar{B}}G^{C\bar{C}} \left( f_{AC}^Df_{BD}^E\eta_{\bar{A}\bar{B}}  + {\rm c.c.}	\right) = 
		\frac{k g_\perp}{2}  \left(
		\begin{array}{ccc}
			0 & g_\perp^2 +g_\parallel^2  & 0 \\
			g_\perp^2 +g_\parallel^2	  & 0 & 0 \\
			0 & 0 & 2g_\perp g_\parallel \\
		\end{array}
		\right)
	\end{equation}
	which coincides with the known answer if the coefficient in front of the action is taken to be $\alpha'=-1/2$. 
	
	\section{Current correlation function}\label{JJJ}
	
	Here we discuss the three-point correlation functions of the currents. 
	For convenience, we recall two-point correlation functions 
\begin{equation}
	\langle J_A(z)J_B(0) \rangle  = -\frac{\eta_{AB}}{z^2},\qquad \langle J_A(z) I^B(0)\rangle = \frac{\delta_{A}^B}{z^2},\qquad \langle I^B(z)I^C(0)\rangle = 0. 
\end{equation}
Here, in principle, $\eta_{AB}$ can be a function of the target fields. 
Recall that the dependence of the target fields dependence is understood as described in \eqref{gb}.
As for the holomorphic fields one can integrate only over the half of the fields. 

Notice that even though the one-point function of the current vanishes $\langle J_A \rangle = 0$, 
the correlator with the non-constant function $F(z) = F(\varphi(z))$ reads as 
\begin{equation}
    \langle J_A(z) F(w) \rangle =\frac{V_A^k \partial_k F}{z-w}.
\end{equation}
The three-point functions can be computed by the following trick. First we identically present 
\begin{equation}
	\langle I^C(x) J_A(z)J_B(w) \rangle = \oint\limits_{C_x}\frac{dy}{y-x} \langle I^C(y) J_A(z)J_B(w) \rangle
\end{equation}
where integration is over a small contour around $x$. Since the correlation function is $O(1/y^2)$ we can deform the contour such that it encircles points $z$ and $w$, inside the contours we can use OPE, which gives us the following 
\begin{equation}\label{IJJ}
	\langle I^C(x) J_A(z)J_B(w) \rangle = -  \oint\limits_{C_z\cup C_w}\frac{dy}{y-x} \langle I^C(y) J_A(z)J_B(w) \rangle = 
	\frac{f_{AB}^C}{(x-z)(x-w)(z-w)}
\end{equation}
The four-point correlation function \eqref{J4} can be obtained in the similar manner. 

It is interesting that the correlation function \eqref{IJJ} is as if currents are primary operators, even though we have not demanded it. 
The \textit{non-primary} effects can be seen on the following correlation function (which luckily is not required for the beta-function computation)
	\begin{equation}
		\langle J_A(x)J_B(y)J_C(z) \rangle  = \frac{\gamma_{ABC}}{(y-z)^2(y-x)} + \frac{\gamma_{ACB}}{(y-z)^2(z-x)} 
		+ \frac{\gamma_{CAB}+\gamma_{CBA}}{(x-y)^2(y-z)} + \frac{\gamma_{BAC}+\gamma_{BCA}}{(x-z)^2(z-y)}
	\end{equation}
	Here 
	\begin{equation}
		\gamma_{ABC} = f_{AB}^D \eta_{DC}  + \partial_m \partial_a V_A^k \partial_k V_B^a V_C^m.
	\end{equation}
		The symmetries of the correlation function are respected due to the following identities 
	\begin{equation}
		\gamma_{CAB}+ \gamma_{CBA} = \gamma_{ABC}+ \gamma_{BAC} = V_C^k \partial_k \eta_{AB}
	\end{equation}
	Notice that contrary to the other current three-point correlation functions \eqref{IJJ} this one 
	does not have a form predicted for $(1,0)$ operators. We see that the form is restores for constant $\eta_{AB}$. 
	Indeed, in this case 
	 $\gamma_{ABC}$ is a completely antisymmetric tensor
	\begin{equation}
		\gamma_{ABC} = \frac{1}{3!} \gamma_{[ABC]}. 
	\end{equation} 
	and
	\begin{equation}
		\langle J_A(x)J_B(y)J_C(z) \rangle  = \frac{\gamma_{ABC}}{(z-y)(x-z)(x-y)}.
	\end{equation}
	
		\section{Second loop in {\em non}-supersymmetric $\mathds{C}P^1$ model is an
		infrared effect}
	
	In this  Appendix we argue that in perturbation theory the second loop in the $\beta$ function of the {\em non}-supersymmetric $\mathds{C}P^1$ model would be absent in the Wilsonean OPE because it is due to an infrared effect. The argument below is from \cite{shU}, page 674, Ex. 6.3.2.

	Let us start from
	supersymmetric  ${\mathcal N} =(2,2)$ $\mathds{C}P^1$  model. For our purposes we will need the following:
	\beqn
	{\mathcal L} &=& G\left[ \partial_\mu \phi^\dagger\, \partial^\mu\phi +i\bar\psi\gamma^\mu \left(
	\partial_\mu + \Gamma \partial_\mu \phi\right) \psi + ...\right]\nonumber\\[2mm]
	&=&G\left[ \partial_\mu \phi^\dagger\, \partial^\mu\phi +i\bar\psi\gamma^\mu \partial_\mu\psi 
	-i\frac{2}{1+\phi^\dagger\phi}\, \,\phi^\dagger\partial_\mu\phi \,\bar\psi\gamma^\mu\psi +...\right]
	\label{k6208}
	\eeqn
	where
	\beq
	G =G_{1\bar1} = \frac{2}{g^2}\,\frac{1}{(1+\phi^\dagger\phi)^2}\,,
	\eeq
	the dots stand for a four-fermion term irrelevant for our present purposes, and $\psi $ is a Dirac spinor.
	
\begin{figure}[h]
				\centerline{\includegraphics[width=6cm]{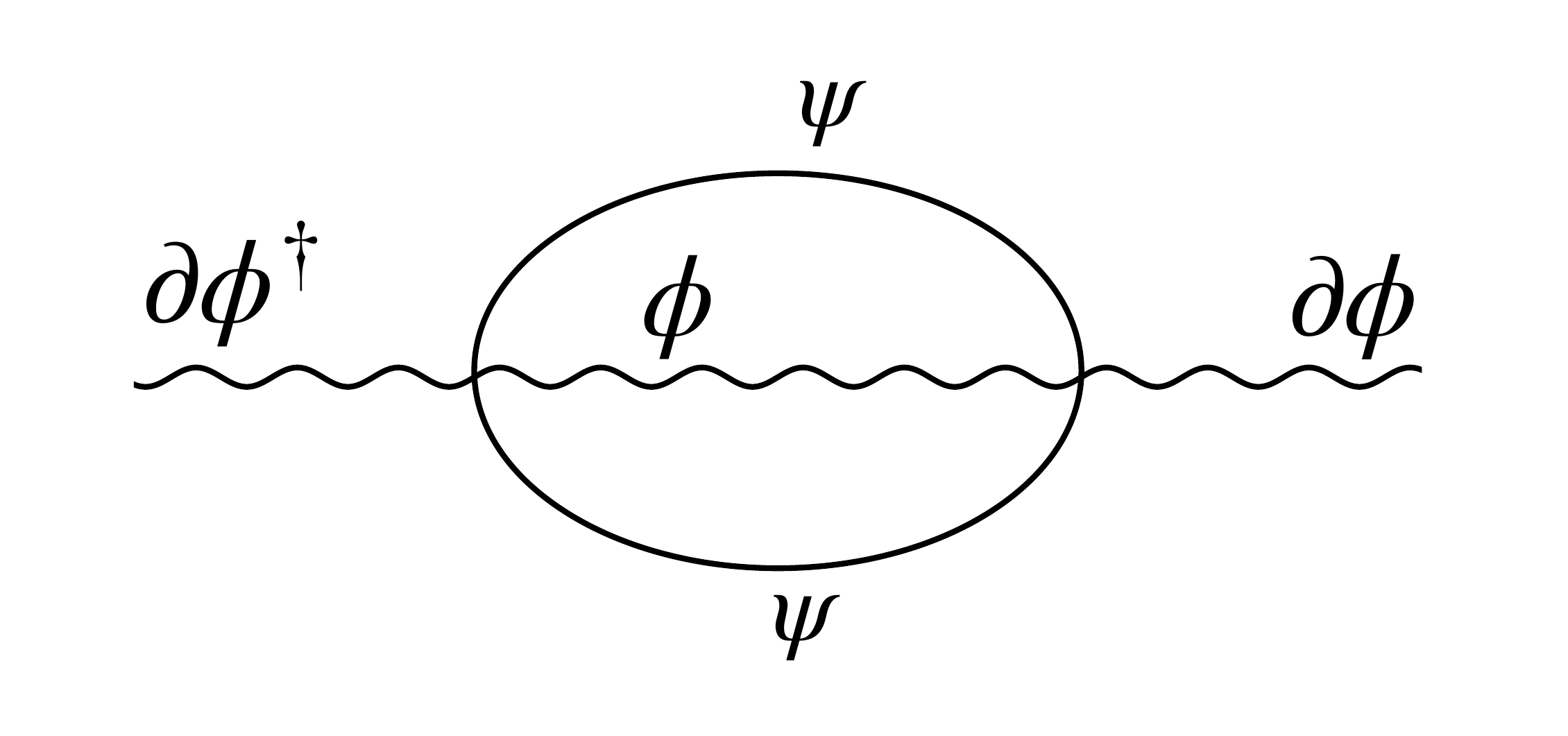}}
				\caption{\small Two-loop contribution in the $\beta$ function of supersymmetric CP(1) model.}
				\label{vtor}
			\end{figure}

	The overall structure of the target space is completely fixed by geometry of $\mathds{C}P^1$. Therefore, to determine renormalization of
	$1/g^2$ in the metric it is sufficient to analyze the vicinity of origin in the target space, i.e. $\phi \approx 0$ (see \cite{shU}, Sec. 6.3.3, page 264). In other words, we can focus on the renormalization of the structure $\partial_\mu \phi^\dagger\, \partial^\mu\phi$  using the last term in (\ref{k6208}) as the interaction vertex, namely,
	\beq
	-i G\, \frac{2}{1+\phi^\dagger\phi}\, \,\phi^\dagger\partial_\mu\phi \,\bar\psi\gamma^\mu\psi \,.
	\label{k6210}
	\eeq
	The only two-loop graph which has exactly the needed structure $\partial_\mu \phi^\dagger\, \partial^\mu\phi$  is 
	depicted in Fig. \ref{vtor}.
	It has three propagators: two fermion and one boson. The denominator in (\ref{k6210}) can be ignored while 
	the $\phi$ fields without derivatives in the numerator  in (\ref{k6210}) and its complex
	conjugated are convoluted into the boson propagator.
	
\begin{figure}[h]
				\centerline{\includegraphics[width=5cm]{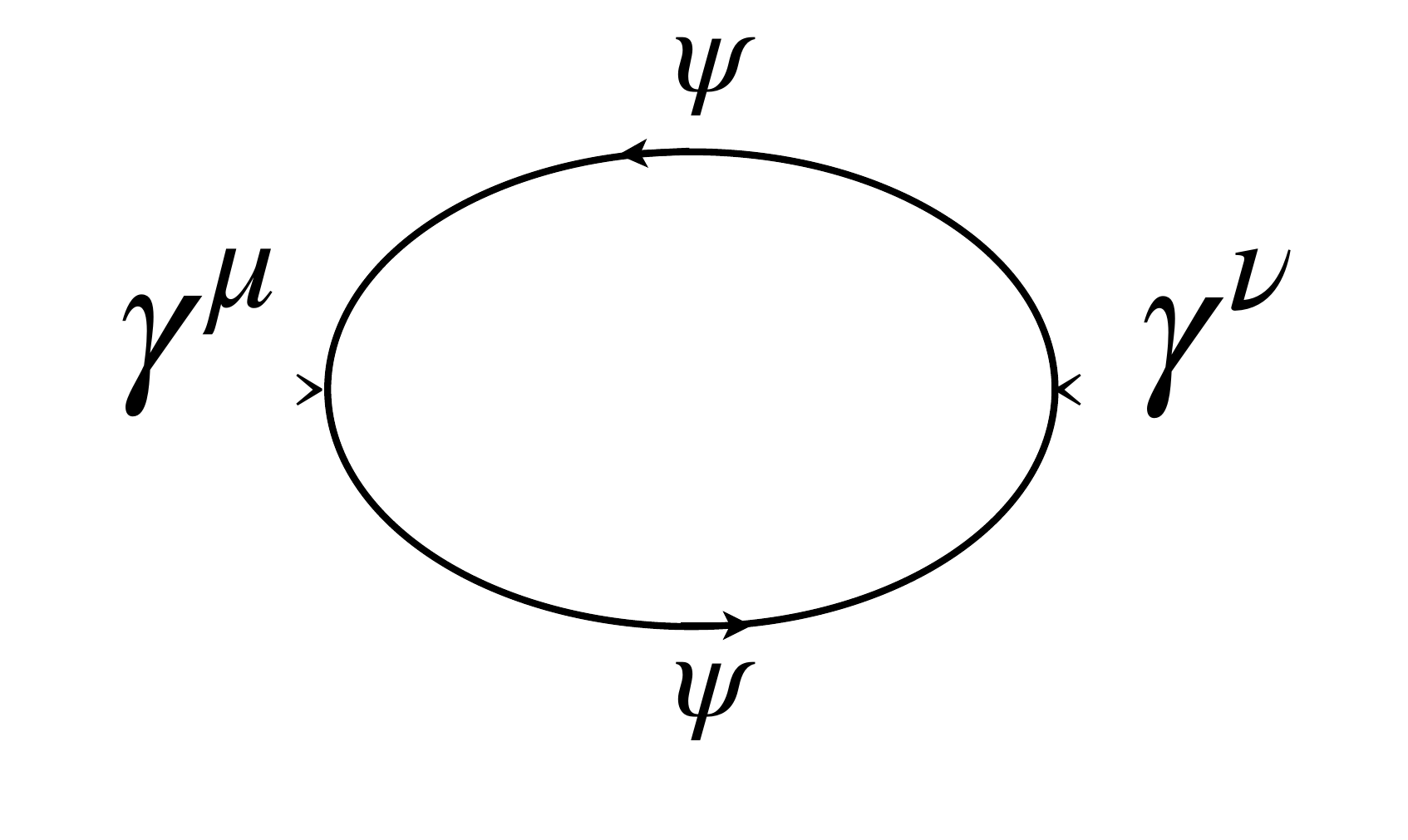}}
				\caption{\small Fermion loop subdiagram.}
				\label{tret}
			\end{figure}

	Next,  we split the calculation in two parts. First we evaluate the fermion loop depicted
	in Fig. \ref{tret}. In $\mathds{C}P^1$ this is the same famous loop which appears in two-dimensional Schwinger model (Fig. \ref{tret}).
	
	\noindent
	Using 
	\beq
	- i\,\frac{1}{2\pi}\,\frac{x_\mu\gamma^\mu}{x^2}\,\, \stackrel{\rm Fourier}{\longleftrightarrow} \,\, \frac{i p_\mu\gamma^\mu}{p^2}
	\eeq
	for the fermion propagator we obtain for the graph in Fig. \ref{tret}
	\beq
	\frac{1}{2\pi^2}\,\frac{1}{x^2}\left(g^{\mu\nu} - \frac{2x^\mu x^\nu }{x^2}\right)\,\, \stackrel{\rm Fourier}{\longleftrightarrow} \,\, 
	\frac{i}{\pi}\left(\frac{p^\mu p^\nu}{p^2} - g^{\mu\nu }\right).
	\label{k6212}
	\eeq
	This is the famous Schwinger infrared effect.
	
	Returning to the diagram in Fig. \ref{vtor} we convince ourselves 
	that the only actions still to be performed are to input the boson propagator
	$i/p^2$ and insert  $(2/g^2)^2$ for two vertices, $(g^2/2)^3$ for three propagators, $-2^2$ reflecting the factor of 2 in (\ref{k6210}) and $-2$ in its complex conjugated, and the overall factor $-i$.
	The final result for the diagram in Fig. \ref{vtor} is
	\beq
	\big(\partial_\mu \phi^\dagger\, \partial^\mu\phi\big) \,\frac{g^2}{2\pi^2}\,\log\frac{M_0}{\mu}
	\eeq
	Since there is no second loop in ${\mathcal N} =(2,2)$ $\mathds{C}P^1$  model, this contribution must be canceled by the sum of all purely bosonic two-loop graphs. Hence, the two-loop renormalization in non-supersymmetric ({\em purely bosonic}) $\mathds{C}P^1$ model is
	\beq
	\delta{\mathcal L}= ... -\big(\partial_\mu \phi^\dagger\, \partial^\mu\phi\big) \,\frac{g^2}{2\pi^2}\,\log\frac{M_0}{\mu} +...
	\label{k6214}
	\eeq
	where the dots stand for irrelevant contributions, $M_0$ is the UV cutoff and $\mu$ is the normalization point. Comparing $\delta{\mathcal L}$ above with $G\big(\partial_\mu \phi^\dagger\, \partial^\mu\phi\big)$
	in (\ref{k6208}) we arrive at the two-loop renormalization of the coupling constant in $\mathds{C}P^1$,
	\beq
	\delta \left(\frac{1}{g^2} \right)_{non\,\,{\rm SUSY\,} \mathds{C}P^1} = -\frac{g^2}{4\pi^2}\,\log\frac{M_0}{\mu}\,.
	\label{k6215}
	\eeq
	This result is certainly well-known in the literature \cite{Grah}. What was not known is the fact that only one-loop in the graph
	is truly UV while the other comes from the IR and, therefore, this contribution would not be counted in the Wilsonean OPE. 
	
	The same conclusion is achieved from the instanton analysis.

 \bibliographystyle{JHEP}
	\bibliography{curbib}
	
\end{document}